\documentclass[twocolumn,floatfix]{revtex4n}
\usepackage{graphics,epsfig,amsfonts,amssymb,amsmath,color}

\begin{document}
\date{\today}
\title{Magnetic interactions in iron superconductors: A review}
\author{E. Bascones}
\email{leni@icmm.csic.es}
\affiliation{Instituto de Ciencia de Materiales de Madrid,
ICMM-CSIC, Cantoblanco, E-28049 Madrid (Spain).}
\author{B. Valenzuela}
\email{belenv@icmm.csic.es}
\affiliation{Instituto de Ciencia de Materiales de Madrid,
ICMM-CSIC, Cantoblanco, E-28049 Madrid (Spain).}
\author{M.J. Calder\'on}
\email{calderon@icmm.csic.es}
\affiliation{Instituto de Ciencia de Materiales de Madrid,
ICMM-CSIC, Cantoblanco, E-28049 Madrid (Spain).}
\begin{abstract}
High temperature superconductivity in iron pnictides and chalcogenides emerges when a magnetic phase is suppressed.  The multi-orbital character and the strength of correlations underlie this complex phenomenology, involving magnetic softness and anisotropies, with Hund's coupling playing an important role.  We review here the different theoretical approaches used to describe the magnetic interactions in these systems. We show that taking into account the orbital degree of freedom allows us to unify in a single phase diagram the main mechanisms proposed to explain the $(\pi,0)$ order in iron pnictides: the nesting-driven, the exchange between localized spins, and the Hund induced magnetic state with orbital differentiation. Comparison of theoretical estimates and experimental results helps locate the Fe superconductors in the phase diagram.  In addition, orbital physics is crucial to address the magnetic softness, the doping dependent properties, and the anisotropies.
\end{abstract}

\maketitle

\section{Introduction}
\label{sec:intro}

\begin{figure}
\leavevmode
\includegraphics[clip,width=0.48\textwidth]{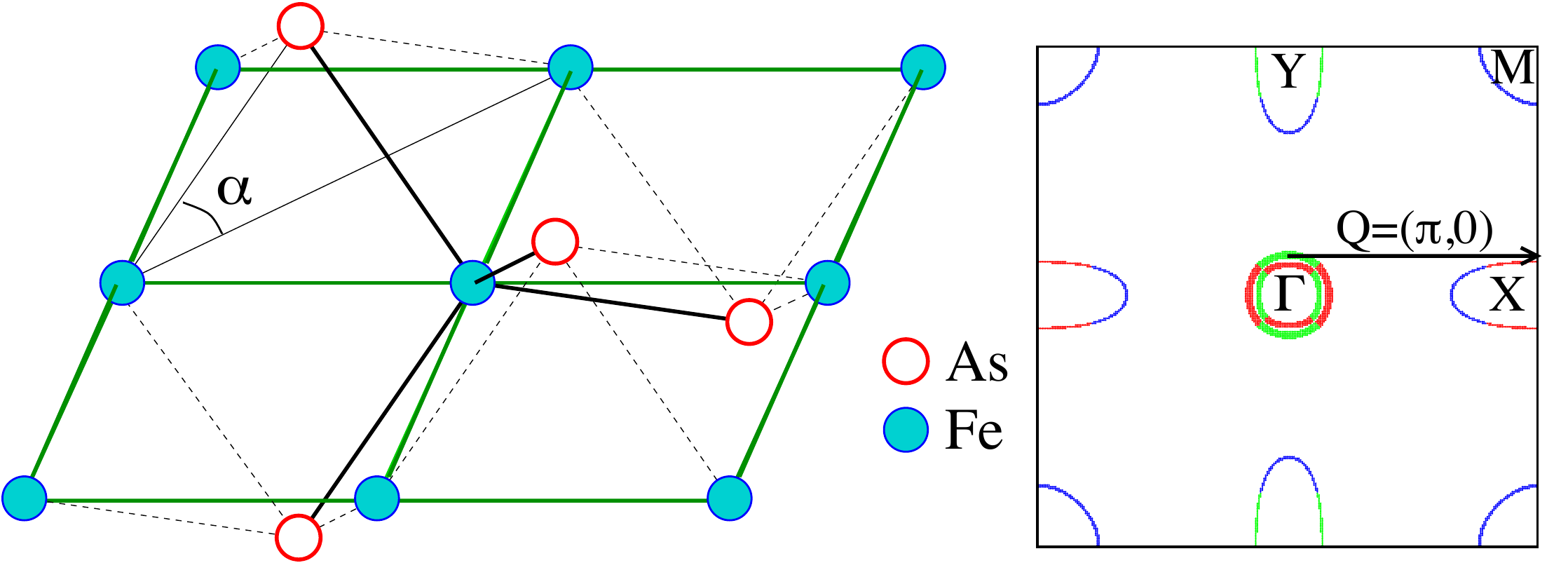}
\caption{(Left) The crystal structure of all the families of Fe superconductors share planes with Fe in a square lattice with a pnictogen or a chalcogen  (here As) located out of plane in a tetrahedral coordination. The relative position of the out-of-plane anions with respect to the plane can be described by the angle $\alpha$. Reproduced from Ref.~\cite{nosotrasprb09}.(Right) Typical Fermi surface defined in the unfolded Brillouin zone (which corresponds to a single Fe unit cell with the $x$ and $y$ directions defined along the Fe nearest neighbours). This Fermi surface has been calculated with the tight-binding proposed in Ref.~\cite{nosotrasprb09} with the $\alpha$ corresponding to a regular tetrahedron. The arrow indicates the wave-vector of the columnar ordering $(\pi,0)$. The pockets at $\Gamma$ and $M$ are hole pockets and the ones at $X$ and $Y$ are electron pockets. The color code refers to the orbital with the highest weight on a particular part of the Fermi surface: Blue for $xy$, red for $zx$ and green for $yz$. The $e_g$ orbitals $3z^2-r^2$ and $x^2-y^2$ are also present at the Fermi surface though with a smaller weight~\cite{nosotrasprb09}.}
\label{fig:lattice-FS} 
\end{figure}

The discovery of iron superconductors~\cite{kamihara06,kamihara08} challenged the uniqueness of cuprates as high T$_c$ superconductors, triggering the search for new materials in previously unexplored directions. Most iron superconductors are magnetic when undoped and the suppression of the magnetism via doping, pressure or isoelectronic substitution is accompanied by the appearance of a superconducting phase~\cite{johnston_review2010,stewartRMP2011,dai_review2015}. There is also a structural phase transition that may coincide with the magnetic transition or occur at a higher temperature. Details may depend on the particular structure or chemical composition but the general trends are robust. All the families of iron superconductors have atomic layers with Fe in a square lattice and a pnictogen (P, As) or a chalcogen (Se, Te) out of plane in tetrahedral coordination~\cite{paglione2010}, see Fig.~\ref{fig:lattice-FS}. 
Understanding the magnetism and electronic correlations in these systems may be the clue for elucidating the yet unknown pairing mechanism for high T$_c$ superconductivity~\cite{dagottonat12,dai_review2015}.

\begin{figure}
\leavevmode
\includegraphics[clip,width=0.95\columnwidth]{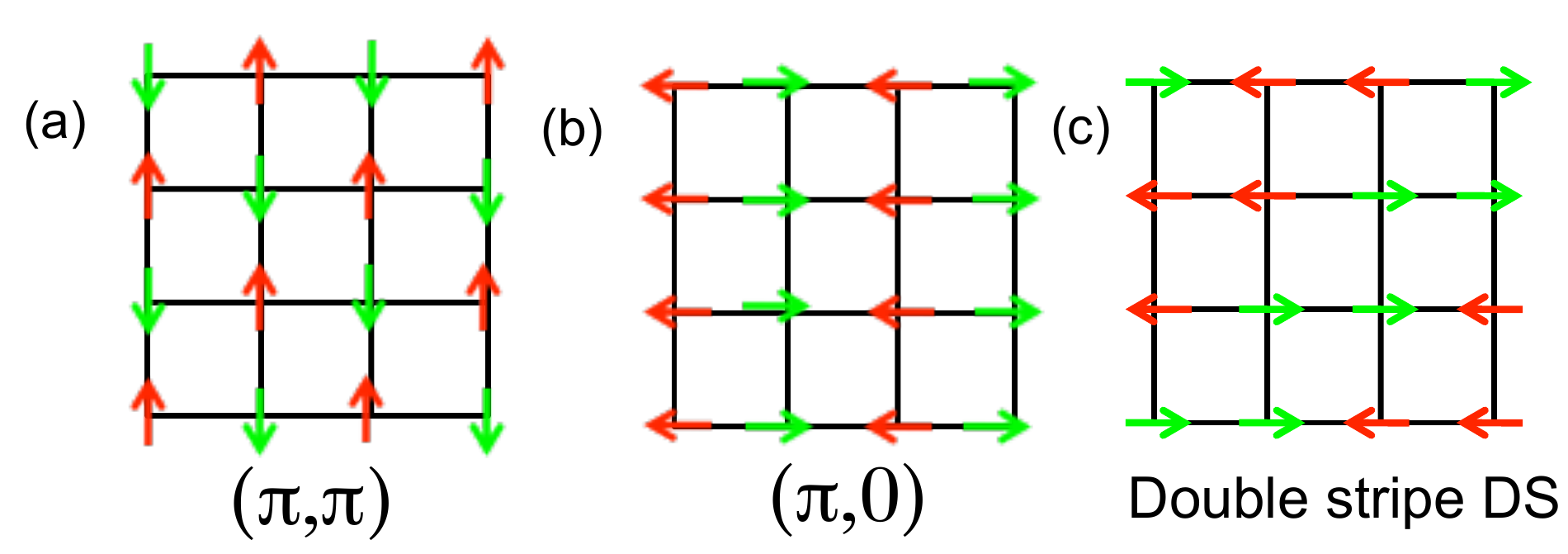}
\caption{Some of the different antiferromagnetic orders within the Fe plane that appear in Fe-based superconductors. Each arrow represents the magnetic moment on an Fe ion. (a) The ${\bf Q}=(\pi,\pi)$ order,  (b) the columnar order ${\bf Q}=(\pi,0)$, the most common order in the Fe-pnictides, and (c) the double stripe order as in FeTe.}
\label{fig:magnetic-orders} 
\end{figure}

Superficially, the iron superconductors phase diagram resembles very much that of the cuprates in the sense that a magnetic phase is replaced by a superconducting one with doping~\cite{dagottoreview}. However, there are some deep differences whose relevance for superconductivity is still to be determined. Undoped cuprates are well known for being Mott insulators and ($\pi,\pi$) order antiferromagnets. On the other hand, undoped Fe superconductors are generally metallic and show different types of magnetic arrangements~\cite{johnston_review2010,dagottonat12}. 
The one that dominates is the columnar ($\pi,0$) one (with axis defined in the Fe-Fe first neighbours direction): antiferromagnetic in the $x$-direction and ferromagnetic in the $y$-direction, see Fig.~\ref{fig:magnetic-orders}. This order (or related magnetic fluctuations) is usually accompanied by a tetragonal to orthorhombic structural transition and in-plane ($x/y$) anisotropies.

The ($\pi,0$) order is found in the undoped iron arsenides, a denomination that includes different chemical compositions and structures --all with FeAs planes. The arsenides mainly comprise the 1111 family, like LaOFeAs, the 122, like BaFe$_2$As$_2$, and the 111, like LiFeAs and NaFeAs. The related phosphides (with FeP planes) are non magnetic and their superconducting T$_c$s are low~\cite{kamihara06}. Most of the arsenides show the columnar order with a small magnetic moment, with typical values under $1 \mu_B$ per Fe, and are metals~\cite{martinelli-issue}. LiFeAs is an exception to the rule as it does not order magnetically, it does not go through a structural transition and it is superconductor~\cite{tappPRB2008}. The iron chalcogenide FeTe has a different magnetic order, the double stripe, namely a double FM column along the diagonals (the Fe-Fe second neighbours direction) and a larger magnetic moment $\sim 2 \mu_B$ per Fe~\cite{baoPRL2009}. On the other hand, bulk FeSe suffers a structural transition at a much higher temperature than the superconducting one but blue static magnetism only arises under pressure~\cite{bendelePRL2010,terashimaarXiv2015}. Related systems arise when some spacer is introduced between the FeSe layers~\cite{blundell-NatMat2013,pachmayrAC2015,pachmayr2014,luNatMat2014,sunIC2015,dongarXiv2015}. The parent compounds of the alkaline Fe selenides A$_y$Fe$_{2-x}$Se$_2$(A=K, Rb,  Cs,  Tl), with Fe vacancies arranged in a particular pattern, have a block antiferromagnetic order with a large magnetic moment of $\sim 3 \mu_B$ per Fe, and are insulators~\cite{weiCPL2011,dagottoreview-selenides}.  Transitions between different magnetic orders have been observed as a function of doping~\cite{MartyPRB2011} and pressure~\cite{bendelePRB2013,monniPRB2013}.

The electronic structure gives us hints to understand the observed metallicity and the differences with cuprates. Models for cuprates usually involve a single orbital. The situation is very different in the iron superconductors. Density Functional Theory (DFT) calculations  show that the density of states around the Fermi level is dominated by the Fe-pnictogen (or chalcogen) planes~\cite{lebegue07,singh08,mazin08}. In undoped compounds, Fe is in a 3d$^6$ valence state and the crystal field is much smaller than the bandwidth. This implies that all the d orbitals may play a role at low energies. The Fermi surface of iron superconductors has both electron and hole pockets at different $k$-points and different orbitals dominate in different parts of the Brillouin zone, see Fig.~\ref{fig:lattice-FS}. Therefore, multi-orbital models are required to describe the Fe superconductors. Although most of these materials are metallic when undoped, correlations are known to play an important role as, comparing ab-initio bands with Angle Resolved Photo Emission Spectroscopy (ARPES) and quantum oscillations measurements, a band renormalisation of $\sim 3$ has been estimated~\cite{luNat2008, terashima2011,sebastian_review}.

The first attempts to understand the phase diagrams of the Fe superconductors came from two opposite approaches: the weak and the strong coupling limits. The weak coupling approach was supported by the metallicity of the parent compounds, the low magnetic moment, and the apparent nesting on the Fermi surface with a ${\bf Q}=(\pi,0)$ wave-vector~\cite{mazin08,dongepl2008}, see Fig.~\ref{fig:lattice-FS}. The strong coupling approach~\cite{yildirim08,si08} was motivated by the observed band renormalisation, the bad metallicity and later by the discovery of families of Fe superconductors whose magnetic order could not be explained by Fermi surface nesting. The multi-orbital character of the Fe superconductors has inspired models in which itinerant and localised electrons coexist~\cite{yin10,lvphillipsprb10,nosotrasprb12,rinconPRL2014}, similarly to other transition metal oxides. Theoretical studies of multi-orbital systems found the possibility of having an orbital selective Mott transition~\cite{shorikov09,demedici2014,yi2013-shen}. But the complexity did not end here, and the effect of correlations in multi-orbital systems has become a research field on its own right. Hund's coupling is one of the main characters of the emergent story, leading to a correlated Hund's metal,~\cite{haule09,fanfarillo2015} with a suppressed but non-zero quasiparticle weight. Different degrees of correlation for the orbitals (orbital differentiation), as observed in ARPES experiments~\cite{yi2013-shen}, adds up to this richness.

An additional complexity arises in connection with the anisotropies observed in these materials. The $(\pi,0)$ magnetic order and the tetragonal to orthorhombic structural transitions break the four-fold rotational symmetry of the lattice. The two transitions usually happen very close in temperature revealing a strong connection between the spin and lattice degrees of freedom. The in-plane resistivity is anisotropic in an unexpected way as it is larger for the shorter ferromagnetic direction. In addition, the magnitude of the anisotropy is largest when the structural distortion and the magnetism are weaker~\cite{mazin10,chuscience2010,yingprl11,chuscience2012,fisherprl14}. Signatures of anisotropy have also been found in other experiments~\cite{Fernandesprl10,meingastprl04,gallaisprl13,sciencedavis10,davisnatphys2013,rosenthalnatphys14,matsudanat12,pengchengdaiprb11,dhitalprl12,pengchengdaiprb13,degiorgi10,uchida2011,degiorgi2012,nakajimaprl12,prozorovnatcomm13,degiorgiprb14,luscience14}. The degeneracy of $yz$ and $zx$ is broken in both the magnetic and non magnetically ordered phases below the structural transition as revealed by ARPES~\cite{shimojima10,shenpnas11,shennjp12} and X-ray experiments~\cite{kim13}. A nematic phase, characterised by the reduction of the point group symmetry from $C_4$ tetragonal to $C_2$ orthorhombic, is found between the magnetic and structural transitions. Nematic transitions are also believed to occur in Sr$_3$Ru$_2$O$_7$ and cuprates~\cite{fradkinannrevcondmatphys10}. 

There is nowadays an intense debate on the origin of the nematicity in the Fe superconductors. Different experiments seem to indicate that it is electronic in origin~\cite{chuscience2010,shenpnas11,yingprl11,chuscience2012,matsudanat12, gallaisprl13,rosenthalnatphys14} but it is difficult to pinpoint between the spin~\cite{nandiprl10,Fernandesreview12} or orbital~\cite{leeyinku09,lvphillipsprb10,kontaniprb11} degrees of freedom due to the spin-orbital entanglement. A recent effective model sensitive to the orbital degree of freedom finds that spin fluctuations generate orbital splitting~\cite{lauraarXiv14}. The possible role of impurities is also under investigation~\cite{uchidaprl13, davisnatphys2013,pengchengdaiprb13,andersenprl14,fisherprl14,hirschfeldarXiv14,breitkreizPRB2014}.
Importantly, orbital or spin fluctuations are also likely candidates for the superconducting pairing glue: magnetic fluctuations are believed to mediate $s^{+-}$-superconductivity~\cite{mazin08,mazinrev09}, and orbital fluctuations are thought to lead to $s^{++}$-superconductivity~\cite{kontaniprl10,kontaniprb11}. 

In this paper we review the magnetic interactions in Fe superconductors with main emphasis on the connection between the orbital degree of freedom and  the correlations, anisotropies and magnetic softness.  In Sec.~\ref{sec:techniques} the different models and techniques used for studying the magnetic properties of the Fe superconductors are introduced.  The following section deals with correlations and the $(\pi,0)$ order and is divided in four subsections:  weak coupling descriptions (Sec.~\ref{subsec:weakcoupling}), strong coupling descriptions (Sec.~\ref{subsec:localised}), ab-initio insights (Sec.~\ref{subsec:abinitio}), and the role of Hund's coupling and the orbital degree of freedom (Sec.~\ref{subsec:hund}). Sec.~\ref{sec:nematic} reviews the origin of anisotropic properties of Fe-superconductors. Sec.~\ref{sec:softness} deals with the current understanding of the observed magnetic softness in these systems. We end in Sec.~\ref{sec:conclusions} with the conclusions.

\section{Models and techniques}
\label{sec:techniques}
Magnetic interactions in iron superconductors have been studied 
using multiple techniques.  Among them,  ab-initio 
methods have played an important role in setting the essential features of the band structure and identifying the magnetic ground state~\cite{singh08,mazin08-2,kuroki08,yin08,yildirimphysc09,johannes09,vildosola08,valenti10,yin10}. DFT methods are single-particle mean field descriptions and deviations from their predictions are expected when many-body effects play an important role in the non-magnetic state. 

A different approach deals with lattice multi-orbital 
systems adding interactions to tight-binding models which 
mimic the bandstructure close to the Fermi surface~\cite{kuroki08,graser09,nosotrasprb09,maprl09,ikedaprb2010,nosotrasprl10,nosotrasprb12}. Models 
from 2 to 8 orbitals per one-Fe unit cell have been used.  
Pnictogen and chalcogen orbitals lie several eV below the 
Fermi level~\cite{nekrasov2008} and many models only include their effect in the 
hopping integrals between Fe-d orbitals.  
Some authors have used only 2 to 4 orbitals per Fe atom. 
However the 5 Fe-orbitals  ($yz$, 
$zx$, $xy$, $3z^2-r^2$ and $x^2-y^2$) 
are required to account in the 
same model for the Fermi surface, the hoppings and the 
orbital fillings, all of which could be relevant in setting the magnetic ground state and correlations. In the following we focus in 
calculations including at least the 5 d-orbitals. We use a one-Fe 
unit cell with $x$ and $y$ axis along the nearest Fe-Fe bonds~\cite{mazin08,patricklee08,nosotrasprb09,weiku2010,kuprl11}.

Typical hamiltonians include the tight-binding, calculated 
within  Slater-Koster~\cite{nosotrasprb09} or by fitting to an ab-initio band structure~\cite{kuroki08,graser09,ikedaprb2010}, the crystal field splitting and local interactions restricted to Fe orbitals: intraorbital $U$, 
interorbital $U'$, Hund's coupling $J_H$ and pair-hopping 
$J'$. 
\begin{eqnarray}
\nonumber
& H &  = \sum_{i,j,\gamma,\beta,\sigma}t^{\gamma,\beta}_{i,j}c^\dagger_{i,\gamma,\sigma}c_{j,\beta,\sigma}+h.c. 
+ U\sum_{j,\gamma}n_{j,\gamma,\uparrow}n_{j,\gamma,\downarrow}
\\ \nonumber & +&  (U'-\frac{J_H}{2})\sum_{j,\gamma>\beta,\sigma,\tilde{\sigma}}n_{j,\gamma,\sigma}n_{j,\beta,\tilde{\sigma}}
-2J_H\sum_{j,\gamma >\beta}\vec{S}_{j,\gamma}\vec{S}_{j,\beta}
\\
& + &  J'\sum_{j,\gamma\neq
  \beta}c^\dagger_{j,\gamma,\uparrow}c^\dagger_{j,\gamma,\downarrow}c_{j,\beta,\downarrow}c_{j,\beta,\uparrow}
+ \sum_{j,\gamma,\sigma}\epsilon_\gamma n_{j,\gamma,\sigma} \,.
\label{eq:hamiltoniano}
\end{eqnarray}
$i,j$ label the Fe sites,  $\sigma$ 
the spin and $\gamma,\beta$ the orbitals. 
Most of the models assume orbital-independent interactions and the 
relations $U'=U-2J_H$, and $J'=J_H$~\cite{castellani78}, 
valid in rotationally invariant systems, leaving only two 
independent parameters, $U$ and $J_H$. Repulsion between 
electrons requires $J_H< U/3$. Interactions have been 
estimated with constrained RPA~\cite{miyake2010}, constrained LDA~\cite{shorikov09} or GW{~\cite{kutepov2010} 
methods. For five-orbital models they give $U \sim 2.5-3$ eV and $J_H \sim 0.35-0.45$ eV in arsenides and $U \sim 3-4$ eV and $J_H \sim 0.45-0.5$ in chalcogenides~\cite{miyake2010}. Models which include $p$ orbitals, as those used in some LDA+DMFT~\cite{haule08,aichhorn2009} or Gutzwiller~\cite{schicklingprl2012} approaches, generally assume non-interacting $p$ orbitals~\cite{yin10,miyake2010} and use larger values of $U \sim 4-5$ eV and $J_H \sim 0.5-0.8$ for d-orbitals, as screening from $p$-orbitals is not included. 

Techniques used to study these multi-orbital models include: Random Phase 
Approximation (RPA)~\cite{raghu2band,kuroki08}, FLEX~\cite{ikedaprb2010}, 
momentum and real-space Hartree-Fock (HF)~\cite{japonesesmf09,nosotrasprl10,tohyamaprb10,dagotto-arpes, nosotrasprb12,nosotrasprb12-2,luoPRB2014}, 
Functional Renormalisation Group (FRG)~\cite{dhleelargo09},  
Dynamical Mean Field Theory (DMFT)~\cite{shorikov09,haule08,haule09,liebsch2010,liebsch2010b,yinNatMater2011,yin11}, Monte Carlo~\cite{imadaPRL2012},  Gutzwiller variational theory~\cite{schicklingprl2011,schicklingprl2012,lanataprb2013,hardy2013-meingast} and 
Slave Spin~\cite{yuprb2012,demedici2014,nosotrasprb14}.  
Different techniques approach the model in weakly or 
strongly correlated limits and can underestimate or 
overestimate the effect of interactions. For this reason and to better understand the underlying physics, the phase diagram as a function of $U$ and 
$J_H$ is often explored.

More simplified descriptions~\cite{chubukov08,korshunov2009,canoprb10,ereminPRB2010,eremin_review,fernandesprb2015} 
consider band models with 
electron pockets at $X=(\pi,0)$ and $Y=(0,\pi)$ and one or 
two hole pockets at $\Gamma=(0,0)$. Bands are 
frequently assumed to be parabolic with no orbital content. Among all 
possible fermionic interactions only density-density and 
pair-hopping between electron and hole pockets, related to 
the spin-density wave instability are kept. These models 
have been treated with Renormalisation Group (RG)~\cite{chubukov08,korshunov2009,fernandesprb2015} or within 
mean field approaches~\cite{canoprb10,ereminPRB2010} for the 
collective magnetic degrees of freedom. 
These low energy models are suitable to deal with itinerant 
electrons in weakly interacting metals, i.e. when 
the ordered states are well described by an instability of 
the Fermi surface and if the orbital content of the Fermi 
surface does not play a relevant role. Very recently a low 
energy model, which only keeps interactions in the spin-channel but retains the orbital character of the bands and 
the information about $U$ and Hund's coupling, has been 
derived~\cite{lauraarXiv14}. 

In the opposite limit, strong coupling approaches~\cite{si08,kivelson08,sachdevprb08,yildirim08,nosotrasprb12} assume that the electrons are localised and 
describe the system with interacting lattice spin models as 
the Heisenberg model 
\begin{equation}
H_{J_1-J_2}={\frac{J_1}{|S|^2}}\sum_{\langle
  i,j\rangle}\vec{S_i}\vec{S_j}+{\frac{J_2}{|S|^2}}\sum_{\langle \langle i,j
  \rangle \rangle}\vec{S_i}\vec{S_j}.
\label{eq:heisenberg}
\end{equation}  
Here $\vec{S}_{i}=\sum_{\beta}\vec{S}_{i,\beta}$ is the atomic moment and
$\langle i,j \rangle$ and $\langle \langle i,j \rangle \rangle $ are 
restricted to first and second nearest 
neighbours respectively. 
The small crystal field splitting in iron 
superconductors suggests $|S|=2$ due to Hund's rule. 
$J_1$ and $J_2$ have been estimated applying perturbation 
theory to multi-orbital models~\cite{haule09,nosotrasprb12} and in ab-initio calculations~\cite{yin08,belaschenkoprb2008,hanprl2009,yildirimphysc09}, 
but in most cases they are assumed to be unknown. In a 
tetragonal state, $J_1$ is equal in $x$ and $y$ directions. 
To fit some neutron measurements, the spin interactions 
to first nearest neighbours are allowed to be anisotropic, i.e.
 $J_{1,a} \neq J_{1,b}$. 
Some measurements and magnetic orderings have been described including in 
the model a $J_3$ interaction to third neighbours~\cite{maprl09}  or a biquadratic term $K (S_iS_j)^2$ to nearest neighbours~\cite{wysocki11,yuprb2012-2}. 
Localised spin models are generally used for insulators. 
They can be used for metals only in the strong coupling 
limit, when interactions place the metal at the verge of a 
Mott transition.

Note that magnetism at finite temperatures and many experimental features require some three-
dimensionality. Nevertheless most of these low-energy, 
localised or multi-orbital, models are two-dimensional.

\section{Electronic correlations and $(\pi,0)$ columnar antiferromagnetism}
\label{sec:pi0}
The nature of  magnetism is  linked to the strength of correlations in the non-magnetic state. Three main pictures have been proposed to explain the $(\pi,0)$ ordering of iron pnictides: an instability of the Fermi surface due to nesting in a weakly correlated state, antiferromagnetic exchange between localised electrons, and a double-exchange like mechanism in a Hund metal state with  orbital differentiation. In the following we  review these approaches comparing with experiment and 
ab-initio calculations. We end the section with a brief discussion.

\subsection{Weak coupling description of magnetism}
\label{subsec:weakcoupling}

The tendency towards a spin density wave instability (SDW) with 
momentum ${\bf Q}=(\pi,0)$  linked to the Fermi surface 
shape was noticed~\cite{mazin08,dongepl2008} before this order was found experimentally~\cite{dongepl2008,cruz08}. Ab-initio 
calculations predict strong nesting between  electron and 
hole pockets connected by ${\bf Q}$, as illustrated in Fig.~\ref{fig:lattice-FS}(b). Nesting 
between idealised circular pockets leads to a divergence in 
the RPA spin susceptibility at ${\bf Q}$. Realistic pockets are elliptical and nesting is not perfect, nevertheless 
a peak in the spin response is obtained~\cite{mazin08,cvetkovic09}.

The antiferromagnetic tendencies emerging from the Fermi surface were early related to an s$^{+-}$ superconducting instability~\cite{mazin08}. FRG calculations showed that the strong 
antiferromagnetic correlation also drives superconducting pairing, Fermi surface 
distortions and orbital current order~\cite{dhleelargo09}.  On the basis of RG calculations, it was argued that magnetic and pairing instabilities are determined by the same 
interband pair hopping~\cite{chubukov08}. When electron and hole pockets are nearly 
identical the SDW instability occurs at a higher T. If the differences in shape or size between the Fermi pockets become more prominent, the 
Cooper instability comes first~\cite{mazin08,chubukov08}.

The suppression of magnetism with doping observed experimentally is expected as 
electron or hole doping changes the size of the Fermi pockets and destroys 
nesting~\cite{zhao08}. This is consistent with the non-monotonic dependence on doping of the minimal 
interaction  for the onset of magnetism $U_{onset}$ obtained in Hartree-Fock calculations 
for a five-orbital model~\cite{nosotrasprl10,nosotrasprb12-2} in which the smallest $U_{onset}$ 
appears at electronic fillings close to that of undoped compounds $n\sim 6$, see Fig.~\ref{fig:magneticphasediag}(d). 
However FLEX calculations question the 
role of nesting as the later is suppressed by self-energy corrections~\cite{ikedaprb2010}. Moreover $U_{onset}> 1$ eV seems too large to claim a nesting induced 
SDW~\cite{nosotrasprl10,nosotrasprb12-2}.  FeTe, with a similar Fermi surface and believed to be more correlated, shows a different magnetic ordering not explicable with nesting arguments, see Sec.~\ref{sec:softness}.

Changes in the size of the electron and hole pockets upon doping modify the momentum at which the RPA spin susceptibility peaks and could produce a change in the ordering moment. 
The tendency towards incommensurate ordering is reduced by Umklapp processes which 
enhance the magnetic instability with momentum ${\bf Q}$, commensurate with the 
reciprocal lattice~\cite{cvetkovic09}. Nevertheless, signatures of incommensurate spin fluctuations appear in the non-magnetic hole-doped compound~\cite{lee_prl2011,daikotliar2013} KFe$_2$As$_2$ and long-range and short-range static incommensurate antiferromagnetic order has been observed in electron-doped BaFe$_2$As$_2$~\cite{pratt_prl2011,luo_prl2012}. While the short-range order was initially interpreted as an evidence of nesting-induced antiferromagnetism, more recent results suggest that it is due to a cluster spin glass phase~\cite{dioguardi_prl2013,lu_prb2014}.

In the reconstructed bandstructure of the magnetic state small Fermi pockets along the $(0,\pi)$ direction have been observed in quantum oscillation and ARPES experiments~\cite{harrison_prb2009,sebastiandirac2011,richard_prl2010}. These pockets are obtained within low-energy effective models~\cite{dunghailee09} and in Hartree-Fock calculations of five-orbital models~\cite{japonesesmf09,nosotrasprl10-2,dagotto-arpes} only if the interactions are not too large. 
In this case the topology of the bandstructure produces nodes in the SDW gap, small Dirac pockets along the $\Gamma-X$  direction and metallicity~\cite{dunghailee09}.

In favour of weak coupling approaches, strong signatures of Hubbard bands have not been observed in X-ray experiments~\cite{bondino_prl2008,cheney_prb2010,manella_review, johnston_review2010}.  ARPES and Quantum oscillations experiments demonstrate the existence of quasiparticles~\cite{luNat2008,yi09,yinNatMater2011,coldea08,terashima2011,sebastian_review}. The shape of the bands is reasonably well described by ab-initio methods~\cite{singh08,vildosola08}. Nevertheless the band mass is considerably enhanced $m^*/m \sim 2-3$ in undoped pnictides~\cite{terashima2011,sebastian_review}.  The low ordered magnetic moment, between $0.09 \mu_B$ in NaFeAs and $1 \mu_B$ in SrFe$_2$As$_2$~\cite{manella_review,dai_review2015}, and the metallicity of the magnetic state 
have been also used in favour of an SDW description of magnetism~\cite{japonesesmf09,dagotto-arpes}. The observed linear increase of the spin susceptibility with temperature~\cite{johnston_review2010} has been explained {with Fermi liquid arguments, as a consequence of antiferromagnetic interactions~\cite{korshunov2009} or due to quasiparticle excitations from a sharp peak in the renormalized spectral function located closely below the Fermi level\cite{citas-anisimov}}. The 
spectral features in the magnetic state of five-orbital models  calculated within Hartree-Fock compare well with optical conductivity, 
photoemission and neutron experiments only if interactions are weak~\cite{nosotrasprb12,japonesesmf09}. Moderate interactions are also 
necessary to reproduce the sign of the resistivity anisotropy 
observed experimentally in calculations with band 
reconstruction as the only source of anisotropy~\cite{nosotrasprl10-2}, see Sec.~\ref{sec:nematic}.

On the other hand, the band reconstruction in NaFeAs observed by ARPES involves bands 
well below the Fermi level~\cite{he_prl2010, johannes09}. 
Inelastic neutron scattering (INS) and X-ray experiments have detected spin waves up to energies above 200 meV~\cite{diallo_prl2009,zhaonatphys09,dagottonat12,inosov_review2015}, 
larger than predicted by weak coupling approaches~\cite{knolle} and estimated total moments  larger  than 
those expected in the weak coupling model~\cite{gretarsson_prb2011,vilmercati_prb2012,
johnston_review2010,dagottonat12,manella_review,daikotliar2013} which do not seem to change much with increasing temperatures up to $T \sim 2 T_N$, well above the N\'eel temperature. This magnetic moment is not only large in systems with magnetic order, but also in compounds which do not order magnetically as LiFeAs or Ba$_{0.6}$K$_{0.4}$Fe$_2$As$_2$~\cite{manella_review}. In a Fermi surface instability picture only the 
bands close to the Fermi level are expected to be affected by 
antiferromagnetism, the spin wave spectral weight is observed only at low energies becoming overdamped by electron-hole pair excitations 
at high frequencies~\cite{tohyamaprb10,andersenarxiv14} and, except for fluctuations effects at temperatures above but close to the transition temperature, the magnetic moments are formed at $T_N$.

\subsection{Antiferromagnetism from localised spins}
\label{subsec:localised}
An opposite point of view to explain the magnetic properties is based on localised spins which interact to first and 
second nearest neighbours with antiferromagnetic exchanges $J_1$ and $J_2$~\cite{si08,kivelson08,sachdevprb08,yildirim08,nosotrasprb12}, see 
Eq.~(\ref{eq:heisenberg}). A $(\pi,0)$ ground state requires $J_2>J_1/2$ as antiferromagnetic ordering with momentum $Q=(\pi, \pi)$ is lowest in energy if $J_2<J_1/2$. For $J_2>J_1/2$ the classical ground state is selected by  $J_2$. 
It consists of two interpenetrating square sublattices with N\'eel order, being the energy independent on the angle between the two sublattices. Columnar order  is selected by quantum or thermal fluctuations through an order by disorder mechanism~\cite{si08,kivelson08,sachdevprb08}. Interactions to first nearest neighbours are frustrated.

The exchange constants $J_1$ and $J_2$ derived at large interactions from multi-orbital models for iron superconductors depend on Hund's coupling and 
on the atomic orbital configuration~\cite{haule09,nosotrasprb12}, see Sec.~\ref{sec:softness}.  The large value of $J_2$ required to fulfill the condition for columnar order $J_2>J_1/2$ is possible due to the role of the As atom in the hopping between second neighbours Fe 
atoms~\cite{yildirim08,nosotrasprb12}. The exchange to first  neighbours $J_1$ includes both direct and indirect 
(via As) contributions.

The presence of magnetic moments above $T_N$ and signatures of well defined spin waves throughout the Brillouin zone are commonly used to justify the local moment picture~\cite{dagottonat12}.  The spin waves measured by neutron scattering can be fitted to a $J_1-J_2$ Heisenberg Hamiltonian but fitting 
parameters are controversial as (i) they can be significantly sensitive to the inclusion of long-range couplings~\cite{ku_arxiv2015}  (ii) the values reported  imply $J_{1a}>>J_{1b}$ with $J_{1b}$ 
slightly ferromagnetic, while in the tetragonal state  $J_{1a}=J_{1b}$. Here $J_{1a}$ and $J_{1b}$ are respectively  the exchange constants along $x$ and $y$ directions~\cite{zhaonatphys09,pengchengdaiprb11}. 
The difference between  $J_{1a}$ and $J_{1b}$ was justified assuming the emergence of orbital ordering $n_{yz}-n_{zx}$, namely different filling of $zx$ and $yz$ orbitals, in the nematic state above $T_{N}$, and a strong impact of this orbital ordering on the exchange constants~\cite{yinleeku10,singh-09}.
Different $J_{1a}$ and $J_{1b}$ have been obtained in ab-initio calculations at low temperatures~\cite{yin08,belaschenkoprb2008,yildirimphysc09}, see below. A spin localised model is generally consistent with orbital fillings not far from  integer. This is at odds with substantial orbital ordering. 
A different  proposal~\cite{wysocki11} to explain the difference between $J_{1a}$ and $J_{1b}$ involves a non-negligible value of the biquadratic coupling $K$ between nearest neighbour spins. 
$K$ induces an Ising degree of freedom and it is behind the selection of $(\pi,0)$ or $(0,\pi)$ order from the classical manifold as well as behind the nematic and structural transitions above $T_N$.  In a columnar ground state the inclusion of $K$ gives $J_{1a} = J_1 + 2 KS^2$ and 
$J_{1b} = J_1 - 2 KS^2$.  
For well localised spins, $K$ is generated by fluctuations~\cite{kivelson08,sachdevprb08} and it is small. Therefore the need of  a large value of $K$ to fit the experiments is believed to imply deviations from the local moment picture~\cite{wysocki11}.

One of the main drawbacks of the localised spin model is the size of the ordered magnetic moment (see Table in Ref.~[\onlinecite{dai_review2015}]). The crystal field arrangement of iron pnictides predicts that the 6 electrons may be accommodated in five d-orbitals with atomic magnetic moment $|S|=2$ (high Hund's coupling) or $|S|=1$ (small Hund's coupling). These atomic moments  are too large compared to the magnetic moment measured by neutron diffraction experiments:  $\mu \sim 1 \mu_B$ in SrFe$_2$As$_2$ and smaller in other compounds~\cite{dai_review2015,inosov_review2015}. 

The early proposal that the frustation of first nearest neighbour interactions could strongly reduce the predicted magnetic moment has not been supported by explicit calculations~\cite{cvetkovic09,kivelson08,carlson2008,thalmeierprb2010}.  The extent
of frustration depends crucially on the ratios of exchange constants and their anisotropies and it is reduced if $J_{1a} \neq J_{1b}$.  The suppression of the magnetic moment is small for $1<|S|<3$ and it decreases with increasing $J_2$~\cite{thalmeierprb2010}.  Unrealistically, low atomic spin and  $|J_2/J_1|$ values very close to $1/2$ would be required to explain the strong reduction of the ordered magnetic moment found in experiment. 

The magnetic moments at high temperatures estimated by integrating the spectral weight in neutron scattering experiments is much smaller than expected for localised electrons~\cite{johnston_review2010,manella_review}. The moments detected by core level photoemission spectroscopy (PES)~\cite{vilmercati_prb2012} and x-ray emission spectroscopy (XES)~\cite{gretarsson_prb2011} are larger than those measured by INS. However the different results of these techniques suggest fast fluctuations of the atomic spin, beyond the localised model~\cite{manella_review}. Consistently, the localised moment approach can neither explain the temperature dependence of the spin excitations~\cite{harrigerprb2012} nor the longitudinal spin excitations detected~\cite{wangprx2013} in BaFe$_2$As$_2$: The spin waves in a localised moment model should be purely transverse spin excitations with moments fluctuating perpendicular to the staggered magnetisation.

\subsection{Insight from ab-initio calculations}
\label{subsec:abinitio}
The $(\pi,0)$ order has been widely studied with DFT methods. Early calculations resulted in different orderings in the 
ground state: proximity to weak ferromagnetism~\cite{singh08,xu-fang2008}, $(\pi,\pi)$ antiferromagnetism~\cite{xu-fang2008,cao-hirschfeld2008} 
and $(\pi,0)$ stripe antiferromagnetism~\cite{dongepl2008,yildirim08}. It was then clarified that $(\pi,0)$ is the magnetic ground state of iron pnictides and that  pseudopotential approaches have to be used with 
special care, only after being checked against all-electron 
methods~\cite{mazin08-2}, because they can fail to identify the correct ground state due to the small differences in energy between the different magnetic orders found in these materials. 
Some calculations were interpreted in favour of a weak-coupling 
origin of magnetism~\cite{zhangprb2010}, but others arrived to 
opposite conclusions~\cite{yildirimphysc09, johannes09}.

The magnetic moment obtained~\cite{mazin08-2} in the $(\pi,0)$ state is   $\sim$2$\mu_B$, much larger than the experimental one 0.35-1$\mu_B$~\cite{dai_review2015}. Such an overestimate of the magnetic moment is rare in DFT.
A proposal to solve this discrepancy involves a high moment ground state with well defined spin density waves but with magnetic twins and antiphase boundaries dynamic on the time scale of the experiment~\cite{mazinnatphys09}. 

The calculations evidence strong magnetoelastic coupling: (i) The moment is very sensitive to the As-z position.  Small displacements of the As change the Fe-moment from 2.0 $\mu_B$  to 0.5 $\mu_B$~\cite{yin08,yildirim08,mazin08-2,belaschenkoprb2008}; and (ii) the atomic positions, generally well described by GGA calculations, show important disagreements with experiments~\cite{mazin08-2,yildirim08-2,yildirimphysc09}. The discrepancies in the crystal structure are smaller in non-magnetic materials while spin polarised calculations are
in agreement with the experimental structure in the non-magnetic state~\cite{mazin08-2,yildirimphysc09}. Magnetic order has to be included to reproduce the experimental 
phonon spectrum~\cite{fukuda2008,yildirimphysc09}.

The sensitivity to the approximation used and to tiny details of the crystal structure was initially  ascribed to the itinerancy of the magnetic ground state which was argued not to be describable in terms of local magnetic moments~\cite{mazin08-2}. The discrepancies between non-magnetic calculations and the experimental structure were later interpreted as an evidence that in experiment there are  magnetic moments at high temperatures, above $T_N$, which cannot be well captured by DFT methods~\cite{yildirimphysc09}.

The value of the exchange constants has been calculated by DFT  using different methods: (i) comparing the energies of the magnetic states and mapping them to a Heisenberg model,  (ii) using the direct spin-flip method, appropriate for high-temperatures~\cite{yildirimphysc09}, and (iii) with linear response perturbation theory, suitable for low temperatures~\cite{yildirimphysc09,hanprl2009,belaschenkoprb2008}. 
The $J_2$ obtained from the magnetic state energy is large, as required to stabilize the stripe ordering in the localised picture.  However, the validity of the method was questioned by a perturbative calculation which found that different magnetic states are stabilised in different atomic orbital configurations~\cite{nosotrasprb12},  see Sec.~\ref{sec:softness}.

Direct spin-flip and linear response perturbation theory provide the dependence of the exchange constants on the distance. Interactions are short-range in local models and long-range when magnetism is driven by nesting of the Fermi surface. In both approaches, the exchange constant along the diagonal direction was found to be antiferromagnetic and very short ranged ~\cite{yildirimphysc09}. Along $x$ and $y$ axis, they decay more slowly~\cite{yildirimphysc09}, as $1/R^3$. Surprisingly, $J_{1b}$ is very different in both methods: it is antiferromagnetic and close to $J_{1a}$ at high temperatures (direct spin-flip method) and ferromagnetic at low temperatures (linear response)~\cite{yildirimphysc09,hanprl2009}. The later result is compatible with neutron experiments. However, different  $J_{1a,b}$ at low and high temperatures is at odds with a Heisenberg description~\cite{yildirimphysc09}. The different magnetic moment at which  DFT calculations converge for different magnetic patterns also points against a Heisenberg description of magnetism~\cite{mazin08-2}. 

LSDA calculations allowed variations on the angle $\theta$ formed by the  magnetic moments on the two Fe interpenetrating antiferromagnetically ordered sublattices selected by the classical solution of the $J_1-J_2$ Heisenberg model~\cite{yaresko2009}. They found a strong dependence of the energy on $\theta$ and attributed it to deviations from a Heisenberg description which could be accounted for by a large biquadratic coupling term $K$, which, as discussed above,  can be behind the different values of $J_{1a}$ and $J_{1b}$. A recent study on several families of Fe-pnictides~\cite{glasbrenner_prb2014} has confirmed that a biquadratic term $K$ is required to fit the ab-initio energies with a spin model. $K$ is of the same order of magnitude as the exchange interaction, it depends on the Fe-Fe and Fe-As distances and it is substantially reduced by hole-doping~\cite{glasbrenner_prb2014}.

Based on an analysis of the LDA single electron energies obtained for  BaFe$_2$As$_2$ and FeTe with different magnetic orderings, Johannes and Mazin argued against the binary choice between superexchange and Fermi surface nesting in favour
of a third mechanism that is neither fully localised
nor fully itinerant~\cite{johannes09}. They concluded that magnetic moments at Fe sites appear due to Hund's rule coupling but that superexchange is not operative. The interactions between the Fe moments are
considerably long-range. However the role played by Fermi surface nesting in setting the magnetic state is small as for both materials the energy gain in the magnetic state  is largely due to the band reconstruction far from the Fermi level~\cite{johannes09}.

\begin{figure*}
\leavevmode
\includegraphics[clip,width=0.999\textwidth]{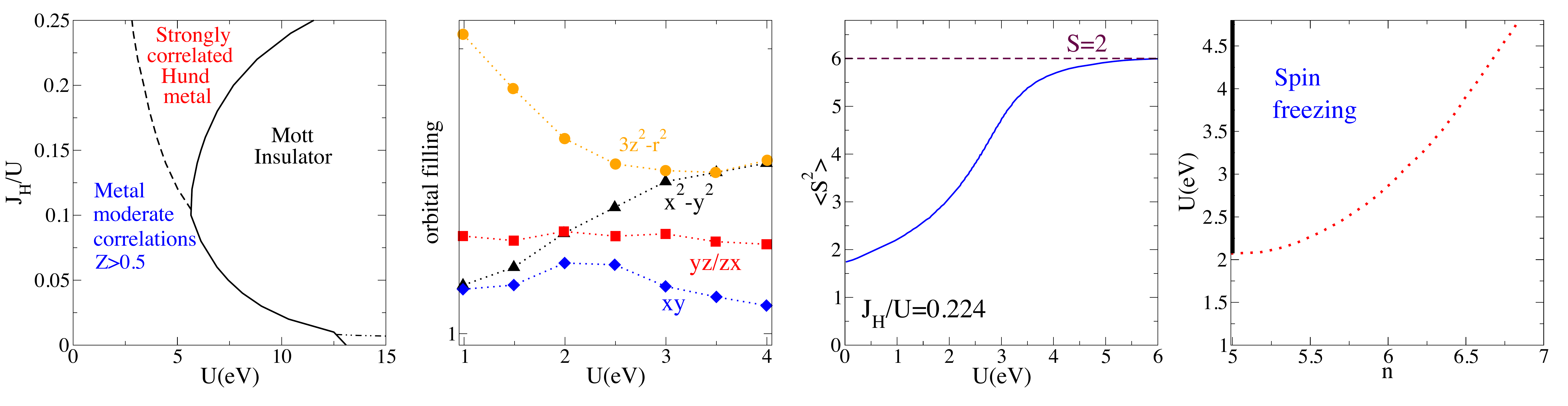}
\caption{
(Colour online)(a) Phase diagram of correlations, as measured by the quasiparticle weight $Z_\gamma$, as a function of the intraorbital interaction $U$ and Hund's coupling $J_H$ calculated with the U(1) Slave Spin technique for undoped superconductors with $n=6$ electrons for a five-orbital model~\cite{graser09} proposed for LaFeAsO, adapted from~\cite{yuprb2012}. Three regions can be differentiated according to the value of the quasiparticle weight: (i) A Mott insulating state at large interactions, (ii) a moderately correlated metal for small $U$ or $J_H$ with $Z_\gamma >0.5$, and (iii) a strongly correlated metal with small quasiparticle weight and strong orbital differentiation (the Hund-metal), see also Figs.~\ref{fig:qpZ}(a) and (b). Mott insulators with $|S|=1$ (low Hund) and $|S|=2$ are separated by a dot-dashed line at large $U$ and small $J_H/U$. (b) Orbital filling versus $U$ calculated with DMFT using full rotational invariant spin interaction for the same tight-binding model in (a), $J_H/U=0.25$ if $U \leq 3$ eV and $J_H=3$ eV for $U>3$ eV, adapted from~\cite{liebsch2010}. (c) Spin fluctuations as a function of $U$ for $J_H/U=0.22$ for a five-orbital model for FeSe. A strong enhancement in the spin fluctuations is observed at the crossover at which the system enters into the Hund metal, adapted from ~\cite{lanataprb2013}. (d) Dependence on doping of the interaction which separates  the weakly correlated state from the spin freezing state at finite temperature, obtained in DMFT~\cite{liebsch2010,liebsch2010b} calculated with the tight-binding model and $J_H$ used in (b). A similar doping dependence is expected for $U^*_{Hund}$ at lower temperatures. With hole-doping $U^*_{Hund}$ approaches the interaction $U_c$ at which the Mott transition happens at half-filling.  } 
\label{fig:hundmetal}
 \end{figure*}

\subsection{The role of Hund's coupling and the orbital degree of freedom}
\label{subsec:hund}
As discussed above, neither weak nor strong coupling approaches are able to explain all the experiments. The renormalisation of the mass, which in a Fermi liquid description is the inverse of the quasiparticle weight, is~$\sim 3$,  emphasizing the importance of correlations.   Aside from a few exceptions~\cite{lauraarXiv14,nosotrasprb12,kemper_prb2011,kuroki09-2} weak and strong coupling approaches ignore the physics emerging from the multi-orbital character of iron superconductors. The simplest generalisation of these models allows the coexistence of itinerant weakly correlated orbitals  and strongly localised ones. However, in the last years  it is becoming clear that Hund's coupling plays a key role on the correlations of these materials and other multi-orbital systems~\cite{shorikov09,haule09,liebsch2010,yuprb2012,nosotrasprb12-2,fanfarillo2015,werner2008,
demedici11-2,demedici11,reviewhund,demediciprl2009}.

  \subsubsection{Correlations induced by Hund's coupling: Hund metals}
  \label{sec:subsubhund}
    DMFT, Slave Spin and Gutzwiller techniques, able to address the renormalisation of the quasiparticle weight, have been used to analyze the correlations induced by Hund's coupling in multi-orbital systems~\cite{werner2008,haule09,fanfarillo2015,reviewhund,demedici11-2,demedici11}. Soon after the discovery of superconductivity in doped LaFeAsO, LDA+DMFT calculations~\cite{shorikov09,haule08,haule09} determined this parent compound to be a correlated metal far from the Mott transition and not well described by either atomic physics or band theory~\cite{haule08}. Hund's coupling was found to be responsible for the electronic mass enhancement~\cite{shorikov09,haule09}. At large $J_H$ the correlated metal, in the following Hund metal, is built from high-spin states with small overlap with single particle states~\cite{haule09,fanfarillo2015,demedici11,reviewhund}, suppressing $Z$. 

 In multi-orbital systems, the Mott transition happens  not only at half-filling but at all atomic integer fillings. The intraorbital interaction $U_c$ for the Mott transition depends on the filling. Hund's coupling modifies $U_c$~\cite{sawatzkyhund,demedici11-2,reviewhund,fanfarillo2015}: (i) At half-filling correlations increase and $U_c$ decreases with $J_H$. (ii) For one electron or one hole occupancies $J_H$ reduces correlations, promotes metallic behaviour and $U_c$ increases.  (iii) Otherwise, $U_c$ is non-monotonous with $J_H$ and induces bad metallicity~\cite{demedici11-2,fanfarillo2015}. 

  The non-monotonic case is relevant to undoped pnictides with 6 electrons in 5 orbitals, Fig.~\ref{fig:hundmetal}(a). With increasing $U$, at intermediate and large $J_H$ the system does not evolve directly from a weakly correlated metal to a Mott insulator~\cite{demedici11,yuprb2012, lanataprb2013,fanfarillo2015,werner2008,liebsch2010}.
  Instead, there is a crossover between the weakly correlated region and a strongly correlated Hund metal with small Z, see Fig.~\ref{fig:hundmetal}(a) and \ref{fig:qpZ}, before the system turns into a Mott insulator~\cite{yuprb2012,lanataprb2013,fanfarillo2015}. The interaction $U^*_{Hund}$ at which the  Hund metallic behaviour appears depends on the number of electrons and on $J_H$ and it can be smaller than the bandwidth $W$~\cite{yuprb2012,lanataprb2013,fanfarillo2015}. In the Hund metal spectral weight is transferred into broad Hubbard bands~\cite{haule08,demedici11-2,liebschpg}.

   The reduction of the quasiparticle weight $Z$ originates in the suppression of atomic configurations which reduce the atomic moment, especially those with double orbital occupancy~\cite{fanfarillo2015}. It is concomitant with the enhancement of spin correlations~\cite{haule09,reviewhund,lanataprb2013,demedici2014, fanfarillo2015} as illustrated in Fig.~\ref{fig:hundmetal}(c), almost saturated in the Hund metal. Hopping processes which do not reduce the magnetic moment are allowed and the system is metallic~\cite{fanfarillo2015,demedici11}. In Fermi liquid theory, the quasiparticle weight is the overlap between the elementary excitations 
of the interacting and the non-interacting systems. 
 The small value of $Z$ in the Hund metal is due to the small overlap between the locally spin polarised atomic states and single-particle states, but it does not necessarily imply strong localisation~\cite{fanfarillo2015} of the atomic charge.

 \begin{figure*}
\leavevmode
\includegraphics[clip,width=0.3\textwidth]{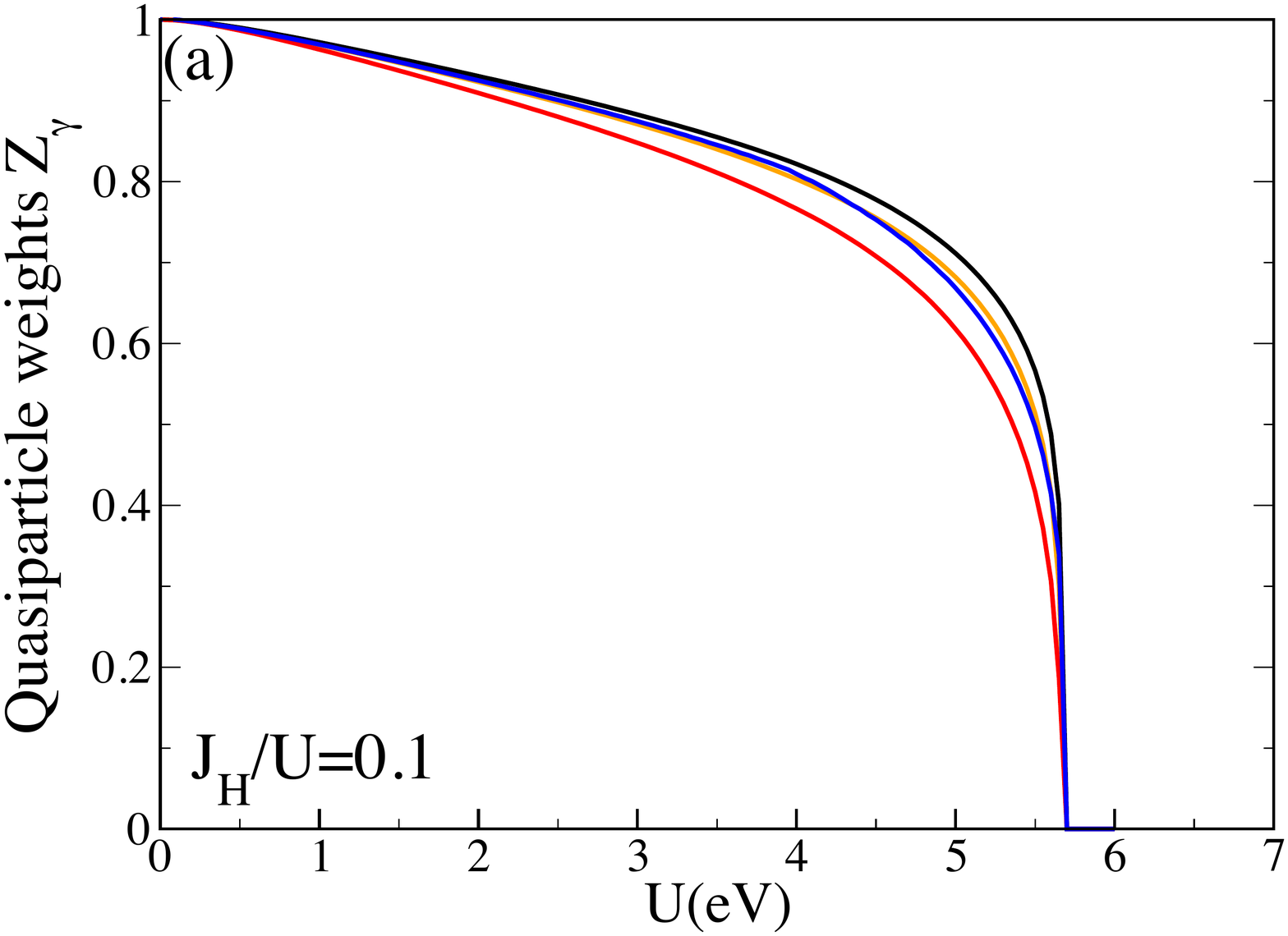}
\includegraphics[clip,width=0.3\textwidth]{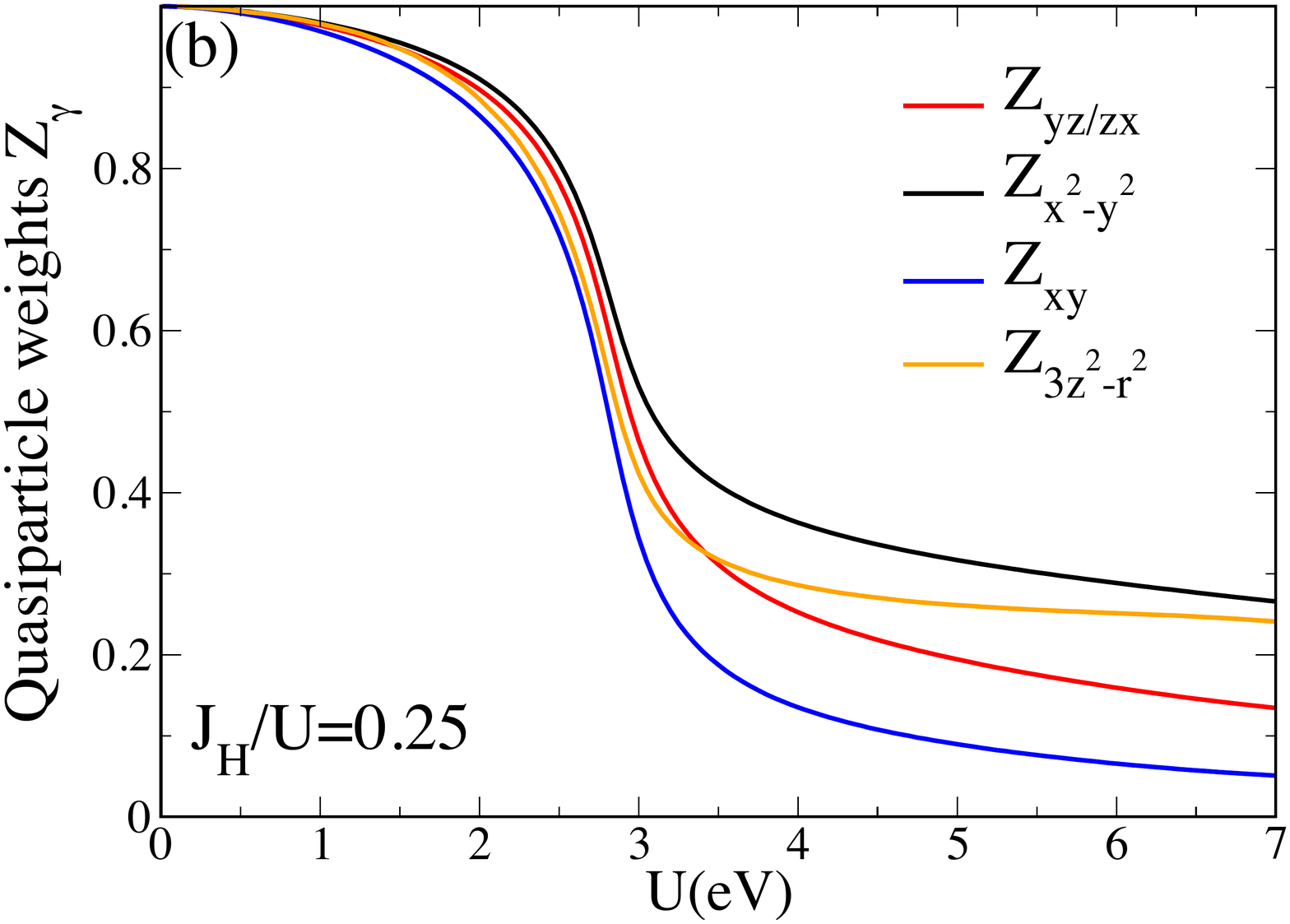}
\includegraphics[clip,width=0.3\textwidth]{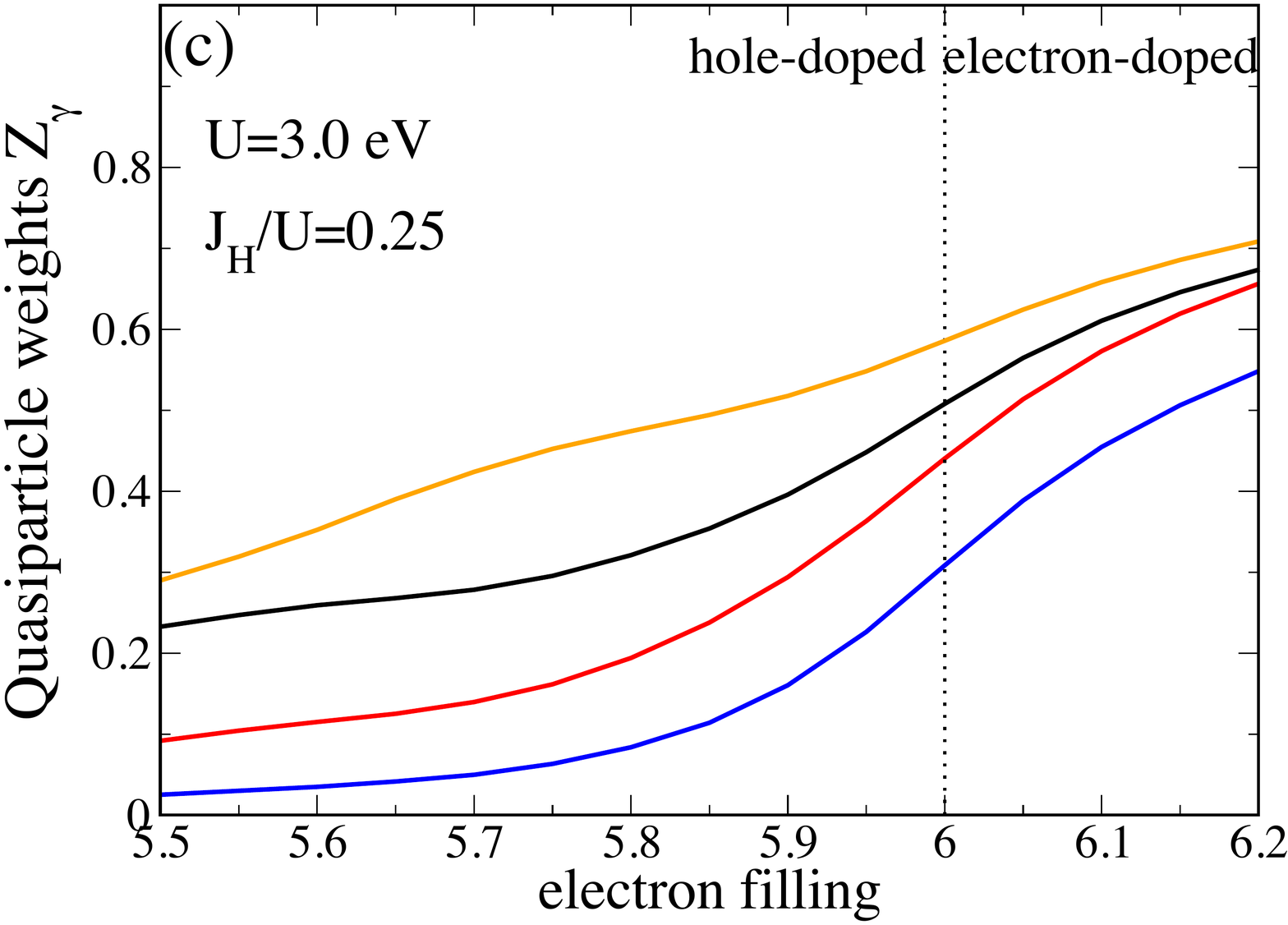}
\caption{
(Colour online)(a) and (b) Orbital-dependent quasiparticle weight $Z_\gamma$ versus intraorbital interaction $U$ calculated with U(1) Slave Spin for $J_H/U=0.10$ and $J_H/U=0.25$ respectively, with $n=6$ electrons as in undoped iron superconductors and the five-orbital model~\cite{graser09} used in Fig.~\ref{fig:hundmetal}(a). In (b) $Z_\gamma$ is strongly suppressed at the crossover between the weakly correlated state and the Hund metal. Orbital differentiation is enhanced in the later. Adapted from~\cite{yuprb2012}.  (c) Quasiparticle weight $Z_\gamma$,  as a function of electronic filling $n$ for the model in (a) calculated with $Z_2$ Slave Spin technique ($U=3$ eV and $J_H/U=0.25$). $Z_\gamma$ decreases with hole-doping as half-filling is approached~\cite{nosotrasprb14}.} 
\label{fig:qpZ}
 \end{figure*}

While the spin correlation functions in the Hund metal paramagnetic state correspond to a large local moment on short time scales, if the electron mobility is high enough, this local moment fluctuates very fast so that  on longer time scales the time-averaged magnetic moment is considerably reduced~\cite{hansmannPRL10}. With increasing temperature the screening time increases and the screened magnetic moment at intermediate time scales becomes closer to its value at short times~\cite{toschi_prb2012}. A crossover towards a highly incoherent metallic state with local frozen moments takes place~\cite{haule09}. 
 A characteristic temperature, the  coherence temperature $T^*$,   can be defined. Below $T^*$ the magnetic moments are screened and Fermi liquid behaviour is observed. Above $T^*$ the spins freeze~\cite{werner2008,liebsch2010,werner2012}. As a consequence, the spin susceptibility evolves from an enhanced Pauli susceptibility (typical of Fermi liquids) at low temperature to a Curie-Weiss behaviour characteristic of localised spins~\cite{haule09}. A Curie behaviour is expected for large screening times~\cite{toschi_prb2012}.
 In the frozen moment state the self-energy shows a non-Fermi liquid power-law dependence~\cite{werner2008,werner2012,kotliarpowerlaw2012} on frequency $\omega^{1/2}$ and the resistivity strongly increases~\cite{haule09}. The coherence temperature $T^*$ becomes smaller with increasing $J_H$~\cite{werner2012}. Screening large local moments is difficult and $T^*$ can be very low.
        
        The value of $U^*_{Hund}$ and of the coherence temperature $T^*$ decrease and the strength of correlations  increases as  the half-filled Mott insulator is approached~\cite{werner2008,liebsch2010,werner2012,demedici2014,nosotrasprb14}. In pnictides this leads to the enhancement of the effective mass and the suppression of the Drude peak~\cite{demedici2014,nosotrasprb14} with hole-doping, see Fig.~\ref{fig:hundmetal}(d) and Fig.~\ref{fig:qpZ}(c). This has led some authors to describe iron superconductors as doped Mott insulators~\cite{liebsch2010,werner2008,demedici2014,nosotrasprb14}, making  connection with the physics of cuprates.  
        
       Based on DMFT calculations, Ishida and Liebsch placed LaOFeAs at the weakly correlated side of this crossover~\cite{liebsch2010,liebsch2010b}. Slave Spin results situate BaFe$_2$As$_2$  around the crossover or just above, into the correlated phase~\cite{demedici2014}. Other LDA+DMFT calculations place most iron superconductors in the Hund metal region~\cite{yin10}.  Chalcogenides are believed to be more correlated than pnictides, especially FeTe and alkaline doped K$_x$Fe$_{2-y}$Se$_2$.  The later compound shows insulating behaviour when vacancies are present~\cite{dagottoreview-selenides}. LDA+DMFT describe FeSe as a Hund correlated metal~\cite{liebsch2010b,aichhorn2010}. Based on Gutzwiller calculations it has been argued than FeSe and FeTe~\cite{lanataprb2013} are Hund metals while using Slave Spin it has been proposed that alkaline doped systems are close the border of the insulating state~\cite{yuprl2013}.

     Experimentally, in most undoped 1111 and 122 FeAs compounds the  mass enhancement estimated from quantum oscillations and ARPES~\cite{terashima2011,sebastian_review,yi09,yinNatMater2011}  is $m^*/m \sim 2-3$. This value  is consistent with the quasiparticle weight $Z$  at the crossover between the two metallic regions or in the Hund metal one~\cite{yuprb2012,fanfarillo2015}, see Fig.~\ref{fig:qpZ}(a). Larger mass enhancement were reported for 111 compounds up to $m^*/m \sim 6$  in NaFeAs from ARPES experiments~\cite{he_prl2010} and $m^*/m \sim 5$ in LiFeAs from quantum oscillations~\cite{putzke_prl2012}.
FeP phosphides are slightly less correlated~\cite{luNat2008} with $m^*/m \sim 1.4-2$ while for FeSe and FeTe the enhancement is between $3.5-7$~\cite{yinNatMater2011}.  The multi-orbital character of iron superconductors make optical experiments difficult to interpret~\cite{benfatto2011,nosotrasprb14} nevertheless the importance of correlations can be also inferred from these measurements~\cite{qazilbash09}.   

The characteristic time scale for the screening of the magnetic moment in iron pnictides was estimated~\cite{toschi_prb2012}  $\sim 10$ fs.  This implies that the magnetic moment detected should depend on the experimental technique, being larger in the case of fast techniques as X-ray, and smaller for techniques which average the magnetic moment in longer time scales such as neutron diffraction, $\mu$SR or NMR~\cite{manella_review}. In INS experiments the time scale is determined by the maximum energy up to which the spectral weight is integrated, but it is nevertheless larger than in the X-ray techniques PES and XAS. In agreement with this prediction, experimentally the values of the magnetic moments estimated by PES measurements are much larger than those obtained by neutron scattering~\cite{vilmercati_prb2012,manella_review,dai_review2015}. Note, that the comparison between PES and neutron measurements is not obvious, as PES determines the total uncorrelated spin while present inelastic neutron scattering measures correlated spin excitations, and thus will underestimate the size of the effective spin when excitations become diffusive~\cite{dai_review2015}.    A temperature dependent screening time could be behind the increase in the magnetic moment with temperature detected in 
Fe$_{1.1}$Te~\cite{zaliznyakPRL2011}

The spin susceptibility at room temperature is enhanced with respect to its bare Pauli spin susceptibility value. In 1111 and 122 compounds the magnetic susceptibility increases with temperature up to, at least, $T \sim 700$~K~\cite{johnston_review2010}. This temperature is probably below the coherence temperature $T^*$ of iron arsenides~\cite{werner2012,toschi_prb2012}. The increasing magnetic susceptibility could be a consequence of the antiferromagnetic correlations~\cite{korshunov2009,johnston_review2010,stewartRMP2011} or due to a sharp peak in the renormalized density of states with an effect similar to that of a van Hove singularity~\cite{citas-anisimov}. On the other hand a crossover is observed in the spin susceptibility of FeSe, it increases with temperature below 180 K and it decreases for larger temperatures~\cite{braithwaite_jphys2009}. The spin susceptibility of FeTe$_{0.92}$ decreases with temperature above the antiferromagnetic transition $T_N \sim 70$ K~\cite{iikubo_jpsj2009}. 

With hole-doping the electronic filling of iron superconductors approaches half-filling (5 electrons in 5 orbitals). Therefore, if undoped iron superconductors are in the weakly correlated region a crossover to the Hund metal state  would be induced by hole-doping. Similarly, if undoped iron superconductors happened to be close to the $n=6$ Mott insulator, they would become Hund metals with electron doping.

       Doping dependent correlations consistent with expectations 
   of Hund's physics have been measured. Different techniques 
   evidence that correlations increase when BaFe$_2$As$_2$ is 
   doped with holes and decrease when it is doped with electrons~\cite{demedici2014}. Specific heat and spin susceptibility 
   measurements indicate that correlations in KFe$_2$As$_2$, 
   strongly hole-doped with $n \sim 5.5$, are stronger than in BaFe$_2$As$_2$ on spite of the smaller cation $K^+$ size which 
   increases the 
   covalency~\cite{hardy2013-meingast,fukazawa_jpsj2011}. Large mass enhancements have 
   been measured in KFe$_2$As$_2$ by quantum oscillations~\cite{terashima2013} and ARPES~\cite{yoshida2012}. The Drude 
   peak in KFe$_2$As$_2$ is strongly suppressed~\cite{uchida2014,uchida2014-2} when compared with that of BaFe$_2$As$_2$ in agreement with theoretical estimates~\cite{nosotrasprb14}. The specific heat enhancement is even larger in  CsFe$_2$As$_2$ and RbFe$_2$As$_2$~\cite{wangprb2013,zhangprb2015}, isovalent to KFe$_2$As$_2$.  Moreover, half-filled Mn-compounds BaMn$_2$As$_2$ and LaMnPO with 5 electrons in the 5 d-orbitals are insulators~\cite{satya_prb2011,simonson_pnas2012} and the strongly electron-doped compound BaCo$_2$As$_2$ with 7 electrons in the 5 d-orbitals shows a small mass enhacement~\cite{xu2013} $m\sim 1.4$. 
     
    Magnetic susceptibility and thermal expansion measurements~\cite{hardy2013-meingast} evidence a temperature induced coherence-incoherence crossover in KFe$_2$As$_2$. At low temperatures ($T<50 K$) the spin susceptibility is nearly independent on temperature while it displays Curie-Weiss behaviour at high temperatures $T>150 K$. Its resistivity is very strongly dependent on temperature with an extremely large residual resistivity ratios RRR$\geq 1000$~\cite{hashimoto_prb2010,hardy2013-meingast}. The superconducting transition is relatively broad, suggesting that this large RRR cannot be only due to reduced disorder but it is a consequence of the temperature induced incoherence~\cite{hardy2013-meingast}. Very large RRR and non-Fermi liquid behaviour have also been measured in CsFe$_2$As$_2$ and RbFe$_2$As$_2$~\cite{wangprb2013,zhangprb2015}. CsFe$_2$As$_2$ shows paramagnetic behaviour: in all the range of temperatures measured the spin susceptibility increases as the temperature decreases~\cite{wangprb2013}.   The temperature dependence of 
 all these measurements is consistent with KFe$_2$As$_2$, CsFe$_2$As$_2$ and 
RbFe$_2$As$_2$ being Hund metals. We note that an alternative explanation for the non-monotonic temperature dependence of the spin susceptibility of KFe$_2$As$_2$ relies on a sharp peak in the density of states below the Fermi level~\cite{anisimov-kfe2as2}. However, to our knowledge, the monotonic increase of the spin susceptibility with decreasing temperature has not been discussed within this single-particle picture.

  \subsubsection{Orbital differentiation in iron superconductors}
  \label{sec:subsubod}

  The orbital-dependent  hoppings and the crystal field splittings render the five Fe-d orbitals inequivalent with consequences on the correlations.  LDA calculations predict for $3z^2-r^2$ the largest occupation in most pnictides, n$_{3z^2-r^2}\sim 1.5-1.6$, while the other orbitals have fillings $n_\gamma \sim 1.0-1.25$~\cite{shorikov09}. Small differences are found between materials~\cite{fillingnote1,fillingnote2}.  With increasing interactions the population of $3z^2-r^2$ calculated with DMFT decreases, that of $x^2-y^2$ increases~\cite{liebsch2010,lanataprb2013,fillingnote1,fillingnote2} and $n_{xy}$ approaches half-filling, see Fig.~\ref{fig:hundmetal}(b). In the Mott insulating state at large $U$, except for very small $J_H$, $x^2-y^2$ is doubly occupied while the other orbitals are half-filled.

  Hund's coupling enhances the differences in filling and quasiparticle weight among the orbitals and can lead to orbital selective Mott transitions (OSMT) ~\cite{demedici2014,shorikov09,yi2013-shen}. In the state with moderate correlations in Fig.~\ref{fig:hundmetal} (a) the differences among the orbital dependent $Z_\gamma$ are small, see Fig.~\ref{fig:qpZ}(a). However, according to DMFT, Gutzwiller and Slave Spin~\cite{liebsch2010,liebsch2010b,yuprb2012,lanataprb2013,
  demedici2014,yuprl2013}, they are substantial in the correlated Hund metal, see Fig.~\ref{fig:qpZ}(b). The t$_{2g}$ orbitals, specially $xy$, become more correlated than the e$_g$ orbitals~\cite{shorikov09,aichhorn2009,liebsch2010,liebsch2010b,demedici2014,
  lanataprb2013,nosotrasprb14,imadaPRL2012,yuprb2012,
  yi2013-shen,yuprl2013,yinNatMater2011,backes_njp2014,ikedaprb2010-2}. Orbital differentiation increases with hole-doping~\cite{demedici2014,nosotrasprb14} and could be behind the sensitivity  to structural details~\cite{yinNatMater2011,lanataprb2013}. 
  
Among the experimental signatures of orbital differentiation are: (i) ARPES and quantum oscillations experiments have reported important differences in orbital correlations in KFe$_2$As$_2$~\cite{yoshida2012,terashima2013}; (ii) The deviations from LDA predictions of the Fermi surface of LiFeAs, LaFePO and LiFeP observed experimentally  have been adscribed to the orbital dependent mass enhancement and the interaction dependent orbital occupancy~\cite{ferberprb2012,ferberprl2012}; (iii) ARPES experiments in 122 chalcogenides have been interpreted in terms of a temperature induced OSMT~\cite{yi2013-shen,yuprl2013}.

The spin excitations measured in neutron experiments~\cite{daikotliar2013,dagottonat12} seem consistent with the presence of both itinerant and localised electrons. Itinerant electrons, sensitive to changes in the Fermi surface with doping, are assumed to be responsible for the low-energy excitations  while localised electrons account for the high-energy spectral weight. The integral of the spectral weight measured with INS, which gives the local fluctuating moment,  is however strongly reduced in the strongly hole-doped KFe$_2$As$_2$, with respect to undoped BaFe$_2$As$_2$~\cite{daikotliar2013}. This result is not easy to understand within a Hund's metal description as a larger atomic moment is expected when half-filling is approached. We are not aware of X-ray estimates of the magnetic moment in this compound.

  The orbital differentiation has consequences on the interpretation of experiments. In particular the suppression of the Drude peak spectral weight in optical conductivity experiments and that of the kinetic energy are not equal, even in a Fermi liquid description ~\cite{nosotrasprb14}. Note that the identification of the Drude peak spectral weight is also affected by interband transitions~\cite{benfatto2011,nosotrasprb14}.

The dependence of interactions on frequency~\cite{werner2012}, non-local correlations beyond single site approximations~\cite{roekeghem_prl2014}, and deviations from the nominal filling not captured by five-orbital models~\cite{aichhorn2010,razzoli2015} have been claimed to have an effect on the correlations in iron superconductors, but they are beyond the scope of this paper. 

  \subsubsection{The role of Hund's coupling and orbital differentiation in the magnetic state of iron pnictides}
  \label{sec:subsubmag}
    Hund's coupling and the orbital degree of freedom have a very significant effect on the
    magnetic state of iron superconductors. 
    Fig.~\ref{fig:magneticphasediag} (a) shows the magnetic phase diagram as a function of the 
    interaction, calculated within $k$-space Hartree-Fock and enforcing $(\pi,0)$ 
    ordering~\cite{nosotrasprl10,nosotrasprb12-2} (see Sec.\ref{sec:softness} for a discussion of other magnetic states). 
\begin{figure*}
\leavevmode
\includegraphics[clip,width=0.25\textwidth]{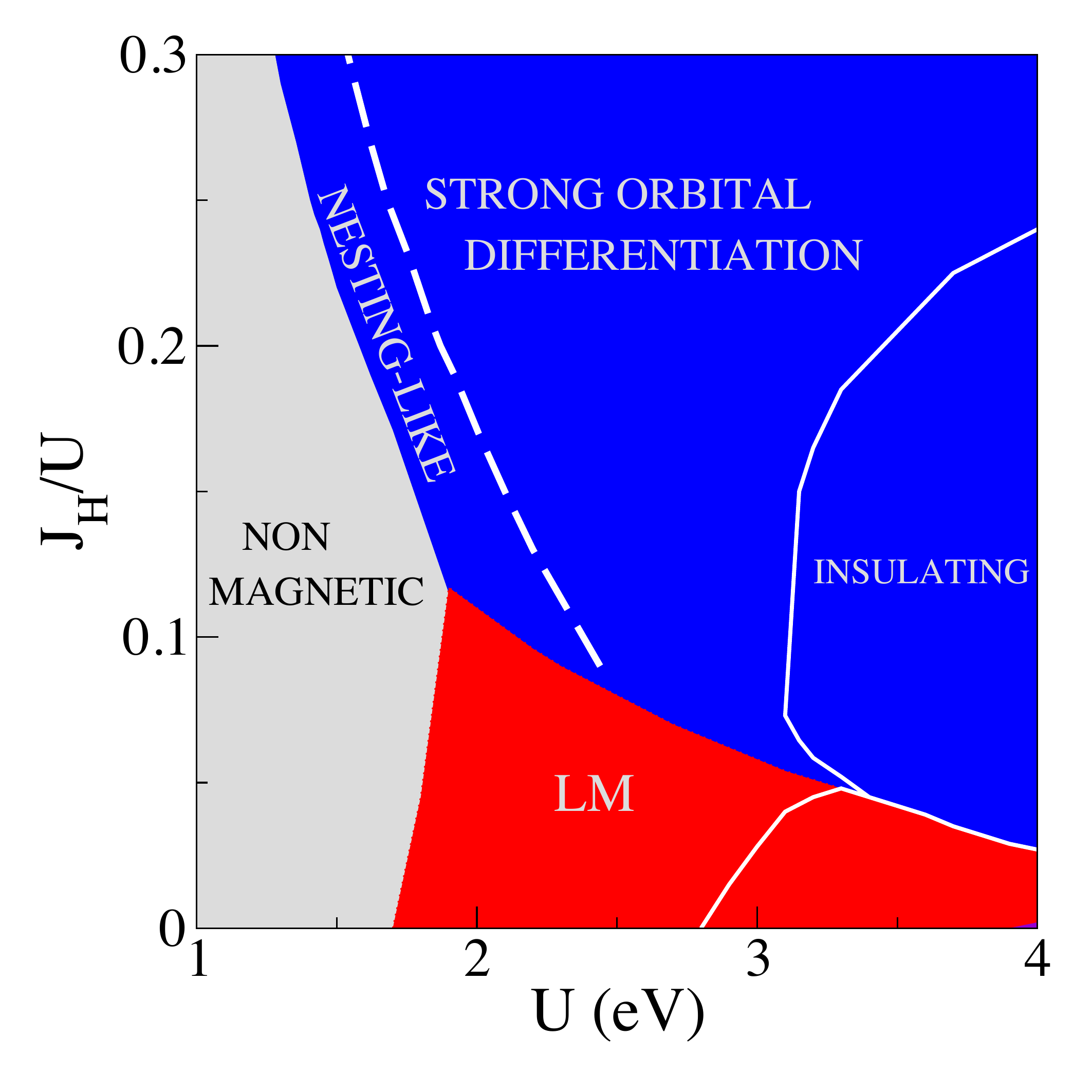}
\includegraphics[clip,width=0.72\textwidth]{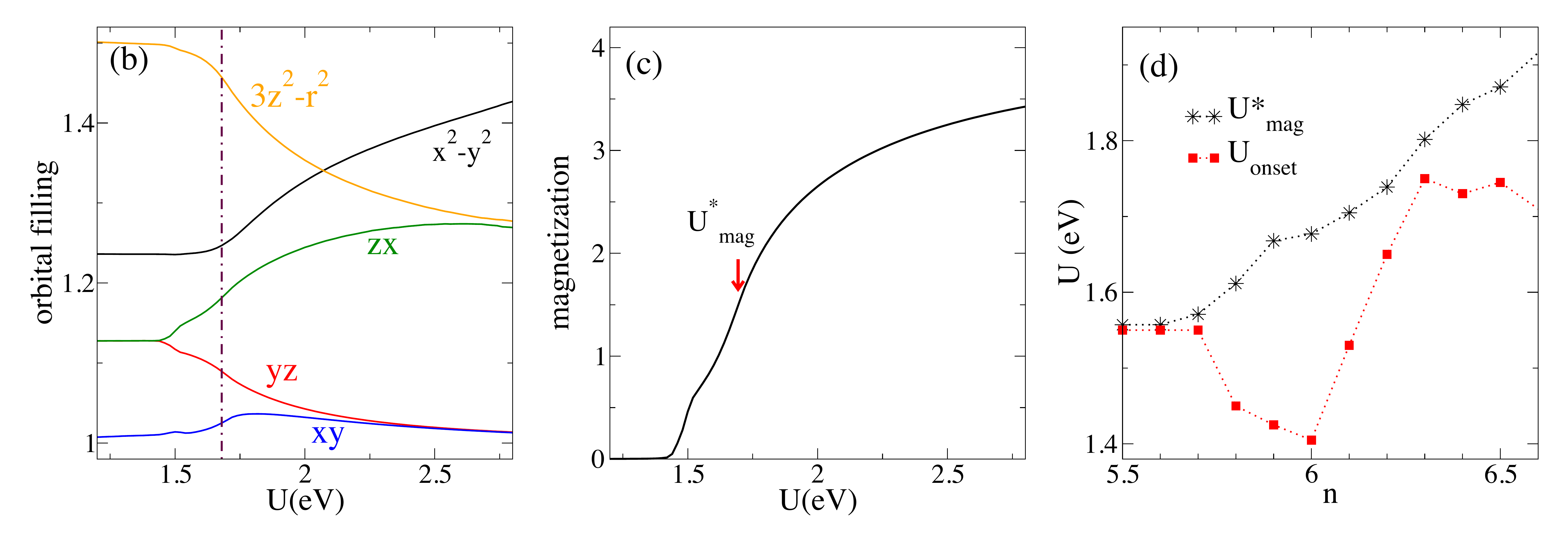}
\caption{
(Colour online)(a) Hartree-Fock $(\pi,0)$ magnetic phase diagram as a function of intraorbital interaction $U$ and Hund's coupling $J_H$ for undoped 
systems with $n=6$ electrons using the five-orbital model proposed in~\cite{nosotrasprb09} (slightly different to that used in Fig.~\ref{fig:hundmetal}), see text. (b) Orbital 
filling in the Hartree-Fock magnetic state as a function of $U$ for $J_H/
U=0.25$. The dot-dashed line marks $U^*_{mag}$, see~\cite{nosotrasprb12-2}. As in the DMFT calculations in Fig.~\ref{fig:hundmetal}(b), with 
interactions $3z^2-r^2$ is emptied, the filling of $x^2-y^2$ increases and $xy$ 
approaches half-filling. The novelty in the magnetic state is the different 
filling of $zx$ and $yz$ consequence of the tetragonal symmetry breaking.  $yz$ 
approaches half-filling and $n_{zx}$ increases. (c) 
Magnetic moment in the Hartree-Fock $(\pi,0)$ antiferromagnetic state as a 
function of $U$ for $J_H/U=0.25$. It shows a strong enhancement at $U^*_{mag}$ 
which resembles the enhancement of spin fluctuations in the non-magnetic state, 
see Fig.~\ref{fig:hundmetal}(c).  (d) In black $U^*_{mag}$, the crossover line 
which separates  the weakly correlated magnetic state and the one with strong 
orbital differentiation, and in red  $U_{onset}$, the interaction at which $
(\pi,0)$ magnetism sets in,  calculated within Hartree-Fock approximation as a function of electronic filling for $J_H/U=0.25$. 
$U_{onset}$ 
is non-monotonous with doping and minimum for $n\sim 6$ as expected in a 
nesting scenario. On the contrary, $U^*_{mag}$ decreases with hole-doping, similarly to 
$U^*$. All figures after ~\cite{nosotrasprb12-2}.
}  
\label{fig:magneticphasediag}
 \end{figure*}
 
NM, in grey, is the non-magnetic state. The phase in red is a low moment (LM) state, with antiparallel magnetic moments, which violates Hund rule. In the blue region, all the magnetic moments point in the same direction and Hund's rule is satisfied. At the right of the solid line the system is insulator, metallic at the left. Insulating phases with $S=1$ and $S=0$, not described here, are found at  $U>4$ eV and  $J_H/U \le 0.02$~\cite{nosotrasprl10}. 

 The LM state, in red, is characterised by a small magnetic moment, see Fig.~\ref{fig:anisotropies}(a), which originates in the partial cancellation of the magnetic 
 moments of different orbitals~\cite{nosotrasprl10}. This state arises from the anisotropy of hopping parameters and magnetic exchanges between {\it orbital} magnetic moments. It emerges as a way to eliminate the
  frustration in the exchange interactions of first nearest neighbours~\cite{nosotrasprl10}. 
This phase was initially proposed to explain the low magnetic moment in 
LaFeAsO~\cite{nosotrasprl10}. In Hartree-Fock it appears for $J_H$ values smaller than expected in real materials~\cite{nosotrasprl10,dagotto-arpes,miyake2010}. With increasing $J_H$, its 
stability is reduced as interatomic and intraatomic exchanges compete. A low 
magnetic moment state violating Hund's rule was also obtained in LDA+U 
calculations for different values of Hund's coupling~\cite{cricchio09,liu11}. A 
Gutzwiller calculation, however, restricted this phase to a small region of 
interaction parameters~\cite{schicklingprl2011}.

Although the values of the interactions  in Fig.~\ref{fig:magneticphasediag}(a) and Fig.~\ref{fig:hundmetal}(a) are not comparable (the effect of interactions is overestimated by Hartree Fock and underestimated by Slave Spin), there are important similarities between the two phase diagrams. The boundary of the insulating blue state in Fig.~\ref{fig:magneticphasediag}(a) shows a non-monotonic dependence on interactions similar to that of the Mott insulator in Fig. \ref{fig:hundmetal}(a). In the insulating blue phase in Fig.~\ref{fig:magneticphasediag}(a), at the right of the solid line, a gap is open at the Fermi level in all the orbitals. $x^2-y^2$ is completely filled and has zero magnetic moment while the other orbitals are half-filled and completely 
spin polarised.  These fillings and orbital polarisations are also found in the Mott insulator in Fig. \ref{fig:hundmetal}(a). Moreover, as discussed in Sec.~\ref{sec:softness} 
the competition with $(\pi,\pi)$ 
ordering found in the insulating state compares very well with expectations based on a mapping to a Heisenberg 
model. Hartree-Fock is a mean field technique and cannot address 
the localisation of electrons, however these similarities suggest a connection between the insulating states Fig.~\ref{fig:magneticphasediag}(a) and Fig.~\ref{fig:hundmetal}(a).  

Similarly to the non-magnetic state, the magnetic phase diagram does not evolve trivially 
with increasing $U$. Changes in the orbital filling are observed, in Fig.~\ref{fig:magneticphasediag}(b).   
$3z^2-r^2$ is partially emptied, $xy$ approaches half-filling and the filling $x^2-y^2$ increases. This tendency is consistent with 
the one observed in the non-magnetic state, see Fig.~\ref{fig:hundmetal}(b). The novelty in the $(\pi,0)$ magnetic state is that the tetragonal symmetry breaking  induces orbital ordering.  The filling of the $zx$ orbital increases  and differs from that 
of $yz$, which approaches half-filling.  Half-filled $xy$ and $yz$ show a gap at the Fermi level, while partially-filled $zx$, $x^2-y^2$ and $3z^2-r^2$ do not, see Ref.~\cite{nosotrasprb12-2}. 

As $U$ is raised, the magnetic moment increases first slowly and then faster at $U^*_{mag}$, see Fig~\ref{fig:magneticphasediag}(c) and Ref.~\cite{nosotrasprb12-2}. This boost coincides with the establishement of strong orbital differentiation leading to two distinct regions in the magnetic phase diagram~\cite{nosotrasprb12-2}. At the left of the crossover $U^*_{mag}$, dashed line  in Fig.~\ref{fig:magneticphasediag}(a), the magnetic moment is small and all the orbitals behave as itinerant, none of them is gapped. At the right of $U^*_{mag}$ the magnetic moment is large and there is strong orbital differentiation: $zx$, $3z^2-r^2$, and $x^2-y^2$ are itinerant while $xy$ and yz are half-filled and gapped~\cite{nosotrasprb12-2}. 

Strong dependence of the magnetic moment on interactions has  also been found in Gutzwiller~\cite{schikling2011}, and  Variational Monte Carlo~\cite{imadaPRL2012} studies of five-orbital models. The enhancement of the magnetic moment at $U^*_{mag}$ is reminiscent of the spin fluctuations at $U^*_{Hund}$. Both crossover interactions $U^*_{Hund}$ and $U^*_{mag}$ have similar dependence on $U$, $J_H$ and electronic filling. They decrease with increasing $J_H$ and with hole-doping while they increase with electron doping. This behaviour is illustrated for fixed $J_H/U$ in Figs.~\ref{fig:magneticphasediag} (a) and (d) and \ref{fig:hundmetal}(a) and (d). We interpret  that in the magnetic state $U^*_{mag}$ is picking up the underlying physics which emerges in the non-magnetic state at $U^*_{Hund}$. As discussed above, the values of $U^*_{Hund}$ and $U^*_{mag}$ in both figures should not be compared as they have been obtained using different techniques.

Note in Fig.~\ref{fig:magneticphasediag}(d) the qualitatively different dependence on doping of $U^*_{mag}$ and $U_{onset}$, the minimal interaction required for the onset of magnetism. While  $U^*_{mag}$ decreases monotonically as the electronic filling is reduced, $U_{onset}$ is non-monotonic with doping and shows a minimum around $n\sim 6$. The doping dependence of $U_{onset}$ and the connection between $U^*_{mag}$ and $U^*_{Hund}$ suggests a nesting-like origin of magnetism, similar to the one described in Sec.~\ref{subsec:weakcoupling} for $U_{onset}<U<U^*_{mag}$. Interestingly, this nesting-like region is not present in the magnetic phase diagram calculated with the tight-binding corresponding to a squashed FeAs tetrahedra, with a Fermi surface characterised by an enhanced ellipticity of the electron pockets and lack of nesting~\cite{nosotrasprb12-2,nosotrasprb13}.

The orbital differentiated region in the $(\pi,0)$ Hartree-Fock phase diagram for $U>U^*_{mag}$ seems connected with the orbital differentiated Hund metal in Fig.~\ref{fig:hundmetal}. 
Hartree-Fock cannot address the renormalisation of the quasiparticle 
weight, nevertheless the observed orbital differentiation suggests a 
description of magnetism in terms of itinerant ($zx$, $3z^2-r^2$,
$x^2-y^2$) and localised  ($xy$ and $yz$) orbitals. This is consistent with 
$3z^2-r^2$ and $x^2-y^2$ being the less correlated  and $xy$ the most 
correlated orbitals in DMFT and Slave Spin calculations in the non-magnetic state, see Sec.~\ref{sec:subsubod}. 
In the non-magnetic state, $zx$ and $yz$ are degenerate and show 
intermediate correlations. The results in the magnetic state suggests  that 
the symmetry breaking and difference in filling between these two orbitals make $yz$ more correlated and $zx$ more 
itinerant~\cite{nosotrasprb12-2}. To our knowledge the orbital differentiation between $zx$ and $yz$ in the magnetic state has not been addressed so far with techniques beyond Hartree-Fock. 

The nature of the magnetism in the orbitally differentiated region differs from the weak and strong coupling approaches discussed before. A double-exchange-like model was introduced~\cite{nosotrasprb12-2} to explain the stability of the $(\pi,0)$ ordering.  Localised $xy$ and $yz$ favour an antiferromagnetic state with $(\pi,0)$ or $(0,\pi)$ momentum depending on parameters, see~\cite{nosotrasprb12-2}. The stability of $(\pi,0)$ magnetism cannot be explained only considering the localised orbitals, see~\cite{nosotrasprb12-2} for a discussion. Itinerant $zx$, $x^2-y^2$ and $3z^2-r^2$, with 4 electrons, promote a ferromagnetic state to gain kinetic energy, specially in the $y$ direction with larger hopping integrals~\cite{nosotrasprb12-2,nosotrasprb09,nosotrasprl10,leeyinku09}. Therefore $(\pi,0)$  ordering  arises as a compromise between the antiferromagnetic tendencies of  the localised orbitals and the ferromagnetic tendencies, larger in the $y$ direction, of the itinerant ones~\cite{nosotrasprb12-2} $zx$, $x^2 - y^2$, and $3z^2 -r^2$. 

A crossover is also observed in the spin wave spectrum of the Hartree-Fock magnetic state calculated including all RPA bubbles and ladder diagrams: the spin excitations evolve from broad low-energy modes at weak interactions to sharply dispersing spin waves prevailing to higher energies at interaction strengths at which $yz$ and $xy$ are half-filled~\cite{andersenarxiv14}. These spin waves, present up to high-energies, are reminiscent of those observed in some neutron experiments.

As discussed in Sec.~\ref{subsec:abinitio}, ab-initio calculations were interpreted in terms of  magnetic moments at the Fe sites which appear due to Hund's rule coupling~\cite{johannes09}, are present in the non-magnetic state~\cite{yildirimphysc09}, and show long range interactions  on spite of the small role played by Fermi surface nesting in setting the magnetic state~\cite{johannes09}. This proposal is compatible with the system being in the Hund metal state. Ab-initio calculations are mean-field techniques similar to Hartree-Fock. A possible interpretation of the large magnetic moments $\mu_B \sim 2$ obtained in ab-initio calculations  and its sensitivity to the As position could be that iron pnictides are in the orbital differentiation region and the  $U^*_{mag}$ line is crossed with small changes in the As position.  Monte Carlo calculations with ab-initio predicted interaction values, in fact,  place different materials at different sides of the $U^*_{mag}$ crossover~\cite{imadaPRL2012}. Undoped and electron-doped 122 materials with larger resistivity in the ferromagnetic direction are not expected to be very deep in the orbital differentiation region as orbital-itinerancy along this direction would favour an opposite resistivity anisotropy\cite{nosotrasprl10-2}, see Sec.~\ref{sec:nematic}.

Following the discussion above, in the Hund metal the local moment fluctuates very fast and the time-average moment is reduced~\cite{hansmannPRL10,toschi_prb2012}. The screened moment is the one which can be magnetically ordered at low temperatures. Temporal spin fluctuations not captured by mean-field methods like Hartree-Fock and DFT would explain the reduced size of the magnetic ordered moment. This is consistent with X-ray measurements which show magnetic moments equal to 2.1$\mu_B$ in SrFe$_2$As$_2$ and  1.3$\mu_B$ in CeFeAsO, respectively larger than their ordered moments 1$\mu_B$ and 0.8$\mu_B$, and than those estimated by INS in other pnictides~\cite{vilmercati_prb2012,manella_review,dai_review2015}.

The spectroscopic properties  of the Hartree-Fock magnetic state, such as optical conductivity or Raman scattering~\cite{japonesesmf09,nosotrasprb13,nosotrasprl10-2}, are mostly sensitive to the size of the magnetic gap which is directly related to the magnetic moment. Therefore the spectrum in the magnetic orbital differentiation region could be significantly different to Hartree-Fock predictions if  temporal fluctuations are included in the calculations and such modifications could alter  the conclusions regarding the region of parameters compatible with experimental results.

In the double-exchange model the value of T$_N$, its interaction and doping dependences are unknown. $n=6$, the filling of undoped pnictides at which $T_N$ is maximum in 122-compounds, does not play any special role in the Hund's metal description.  It is not clear whether $T_N$ should follow  the same dependence on doping as the strength of correlations as with hole-doping $zx$ is emptied losing its itinerant character. Moreover, in single band models, $T_N$ depends non-monotonically on interactions. T$_{N}$ can be also influenced by  the competition with other magnetic orderings, in particular with (i) $(\pi,\pi)$,  observed in the chromium-doped compound and energetically favourable with hole-doping, or with (ii) double stripe or staggered-dimer 
orderings~\cite{nosotrasprb12,valenti2015,luoPRB2014}, found in models for iron pnictides and chalcogenides, see Sec.~\ref{sec:softness}. Studying $T_N$ as a function of interactions and doping, with techniques beyond mean-field, would be very useful. This analysis should, nevertheless, use three-dimensional models.

\subsection{Discussion}

 Previous discussion leaves us without arguments to approach $(\pi,0)$ magnetism and correlations in iron arsenides by considering only local moments:  The materials are metallic, techniques which sample the system in different time scales report different magnitudes of the magnetic moment,  and the ordered moments are small and cannot be justified on the basis of frustration.  Moreover, longitudinal spin excitations have been detected, and the parameters  used to fit the spin waves involve strong anisotropies, inconsistent with the almost commensurate fillings expected in a localised approach, or a large biquadratic term typical of systems which deviate from localisation, as also found in ab-initio calculations.

On the other hand, there are clear signatures of strong correlations and of the role played by Hund's coupling in iron pnictides. The electronic bands are strongly renormalized. The mass enhancements reported are comparable to those found in the Hund metal state in Fig.~\ref{fig:qpZ} (b) and increase with hole doping. Orbital differentiation, i.e. differences in the correlation strength of different orbitals, is observed and becomes more pronounced in hole-doped systems.  There is evidence of magnetic moments far above the N\'eel temperature. Their value is larger than the one measured by neutron diffraction in the ordered state. These moments are  present not only in systems which order magnetically at low temperatures but also in systems which do not order.  The presence of magnetic moments at high temperatures  is in agreement with the need of including magnetism to reproduce the crystal structure and the phonon spectrum. 

Deep in the Hund metal region a Curie-like temperature spin susceptibility originating from a small coherent temperature $T^*$ and larger conductivity in the ferromagnetic direction are expected. Such dependence is not observed in  the undoped systems but it has been found in hole-doped CsFe$_2$As$_2$ in the whole range of temperature measured and in KFe$_2$As$_2$ above 150 K. The anisotropy of the resistivity agrees with the experimental one only for moderate interactions, see below. 

Therefore, our conclusion is that in undoped FeAs systems there is evidence of the formation of atomic moments due to Hund coupling, but the systems are not deep in the Hund metal state, i.e.  undoped arsenides should be placed around $U^*_{Hund}$,  the crossover between the Hund metal and the weak coupling regions. Hole-doping moves them deeper into the Hund metal state. This conclusion is consistent with theoretical estimates which place undoped iron pnictides around the crossover  $U^*_{Hund}$ in the non-magnetic state~\cite{liebsch2010,haule09,yinNatMater2011,werner2012,demedici2014} and around $U^*_{mag}$ in the magnetic state~\cite{imadaPRL2012}. It is also consistent with the strong sensitivity of the magnetic moment to the As position found in ab-initio calculations.  Note that while this crossover is relatively sharp in the slave spin calculations plotted in  Fig.~\ref{fig:qpZ} and computed including only density-density interactions~\cite{yuprb2012,demedici2014,fanfarillo2015}, DMFT calculations which treat the rotationally invariant Hund's interaction find a smoother crossover, see Ref.~\cite{liebsch2010}.   

Experimental evidence and theoretical calculations show that chalcogenides are more correlated. In particular, the spin susceptibility of FeSe and FeTe decreases with increasing temperature, respectively above 180 K and 70 K. Mass enhancements are also larger and an anomalous temperature dependence of the magnetic moment was found in FeTe$_{1.1}$. Moreover the magnetic order in FeTe differs from that  predicted by Fermi surface arguments. In our view, 11 compounds can be classified as Hund metals. The high-moment found in alkaline-doped chalcogenides is expected only deep in the Hund metal region or very close to the Mott insulating state. 

Finally we note that the origin of $(\pi,0)$ magnetism is different at both sides of $U^*_{mag}$:  nesting-like for smaller $U$ and double-exchange for larger $U$. To our knowledge, there are no studies of the interplay between these two mechanisms at the boundary $U^*_{mag}$. Whether one of them dominates, they cooperate or compete is not clear to us.

\section{Anisotropy and the spin-orbital-lattice entanglement}
\label{sec:nematic}
\begin{figure*}
\includegraphics[clip,width=1.0\textwidth]{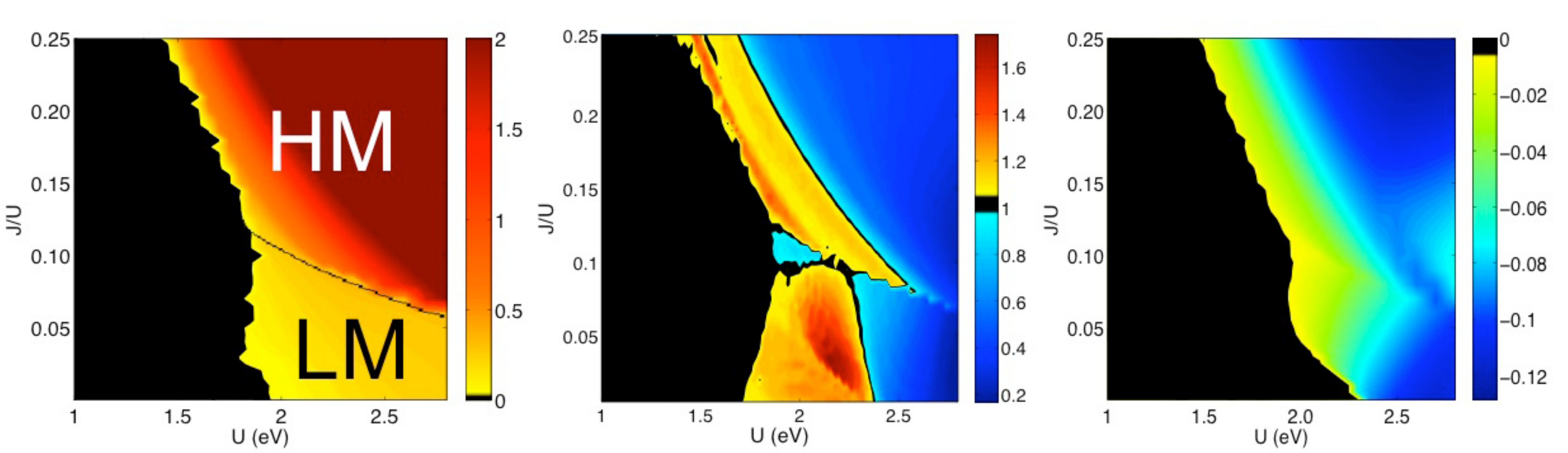}
\caption{ (Color online) (a) $U$ vs $J/U$ magnetic phase diagram superposed to 
the magnetisation showing the non-magnetic (black) and magnetic phases (LM) and (HM). LM stands for low magnetic moment with violation of Hund's rule and HM stands for parallel magnetic moments. The magnetic moment reaches values as high as 3.4 $\mu_B$. The color scale 
emphasizes the region with magnetic moments  smaller than 2 $\mu_B$.  
(b) Drude ratio $D_x/D_y$. The experimental anisotropy corresponds to $D_x/D_y>1$.  Largest values of $D_x/D_y>1$ occur at small 
values of the magnetisation. 
(c) Orbital ordering $n_{yz} - n_{zx}$. The orbital ordering anticorrelates with $D_x/D_y>1$. The figure is reproduced from ~\cite{nosotrasprl10-2}. } 
\label{fig:anisotropies}
\end{figure*}

Besides the discussion on the role of correlations in the columnar antiferromagnetic ordering, another ongoing debate is the connection of this ordering with the nematic state. ${\bf Q} = (\pi$, 0) columnar ordering is linked to the nematic state since it breaks the $C_4$ symmetry of the lattice down to $C_2$ symmetry. This symmetry is also broken by the structural transition and by orbital ordering. Consequently, nematicity could have an orbital, spin or lattice origin and there is a strong debate about its mechanism~\cite{fernandesnatphys14, hirschfeldnatphys14}. The difficulty in deciding between scenarios stems in that they all break the same symmetry, 
which implies that as soon as one symmetry is broken in one channel it will be also broken in the other two channels. Different experiments seem to indicate 
that nematicity is electronic in origin~\cite{chuscience2010,shenpnas11,yingprl11,chuscience2012,matsudanat12, gallaisprl13,rosenthalnatphys14} leaving as candidates the orbital 
and spin degrees of freedom. In the orbital driven nematicity scenario~\cite{leeyinku09,lvphillipsprb10,kontaniprb11} orbital fluctuations provide the pairing mechanism for superconductivity in contrast to the spin fluctuation pairing mechanism for the spin driven nematicity~\cite{nandiprl10,Fernandesreview12}. 

Experimentally, the resistivity is smaller for the antiferromagnetic direction than in the ferromagnetic direction~\cite{mazin10,chuscience2010,yingprl11,chuscience2012,fisherprl14}. Counterintuitively, this resistivity anisotropy grows with doping and is largest close to the border of the antiferromagnetic phase when the magnetism and the structural distortion are weaker.  The opposite resistivity anisotropy has been found in a hole-doped material~\cite{prozorovnatcomm13}. Anisotropy has also been measured in optical conductivity ~\cite{degiorgi10,uchida2011,degiorgi2012,nakajimaprl12,degiorgiprb14,mirriprb14}, elastic shear modulus~\cite{Fernandesprl10,meingastprl04},
Raman~\cite{gallaisprl13}, scanning tunnelling microscopy~\cite{sciencedavis10,davisnatphys2013,rosenthalnatphys14}, magnetic torque~\cite{matsudanat12} and inelastic neutron 
scattering experiments~\cite{pengchengdaiprb11,dhitalprl12,pengchengdaiprb13,luscience14}. Signatures of orbital-dependent reconstruction of electronic structure in the magnetic and non-magnetic states are found in ARPES experiments~\cite{shimojima10,shenpnas11,shennjp12,fengprb12,coldeaarxiv15} and in X-ray absortion spectroscopy~\cite{kim13}. 

In addition, there is no agreement about the relevance of impurity scattering in the $x/y$ 
resistivity anisotropy~\cite{uchidaprl13,davisnatphys2013,pengchengdaiprb13,andersenprl14,fisherprl14,hirschfeldarXiv14,breitkreizPRB2014,uchidaarxiv15}. Upon annealing, Ishida et al.~\cite{uchidaprl13} find that the resistivity anisotropy of BaFe$_2$As$_2$ nearly vanishes while significant anisotropy remains in Co-doped compounds. Similar results are obtained in Ba(Fe$_{1−-x}$Ru$_x$)$_2$As$_2$~\cite{uchidaarxiv15}. In accordance with these experiments scanning tunneling microscopy~\cite{sciencedavis10,davisnatphys2013,rosenthalnatphys14} report anisotropic spatial extended defects, called nematogens, at low temperature.  However, Kuo and Fisher in Ref.~\cite{fisherprl14} find that the strain induced resistivity anisotropy in BaFe$_2$As$_2$, Ba(Fe$_{1-x}$Co$_x$)$_2$As$_2$ and Ba(Fe$_{1-x}$Ni$_x$)$_2$As$_2$ is independent of disorder in the tetragonal phase and it is appreciable in the magnetic state of BaFe$_2$As$_2$. Furthermore, the nematic susceptibility shows clean critical exponents~\cite{chuscience2012}, suggesting irrelevance of disorder. Theoretically, nematogens around impurities have been found in models of iron pnictides driven by orbital~\cite{kontanissc12} or spin correlations~\cite{gastiasoroprb14}. On the other hand, in calculations in the magnetic phase of multi-orbital models the role of the topology and morphology of the Fermi surface in the resistivity anisotropy has been emphasised without~\cite{nosotrasprl10-2} and with impurities~\cite{sugimotoprb14}. 

Hints about the strong interrelation between the spin and orbital degrees of freedom come from first principle calculations and mean field approaches in multi-orbital 
Hamiltonians. It has been found that orbital ordering is present when the $(\pi,0)$ AF 
sets in in realistic models for iron pnictides~\cite{leeyinku09, yu09, nosotrasprl10, nosotrasprl10-2,daghofer10,yinpickettprb10,yin11}. However, to the best of our knowledge, the spontaneous generation of orbital ordering has not been found without magnetism in these type of models.
To study the consequences of orbital ordering in the nematic phase a phenomenological crystal field term breaking $yz/zx$ degeneracy has to be included in the Hamiltonian 
(see for example ~\cite{lvphillipsprb10,schmalianprb12,breitkreizPRB2014,andersenarXiv15}).

Analysis of the Drude weight in multi-orbital models has helped clarify the mechanism of the resistivity anisotropy.
In ~\cite{nosotrasprl10-2} the ratio of the Drude weights $D_x/D_y$ is calculated in the magnetic state assuming an isotropic scattering rate. Fig.~\ref{fig:anisotropies}(a) compares the magnetic moment (a), the $D_x/D_y$ ratio (b) and the orbital order $n_{yz}-n_{zx}$  (c) in the $U$ vs $J_H$ phase space. LM and HM in Fig.~\ref{fig:anisotropies}(a) stand for the phases which violate and fulfill Hund's coupling, respectively, see the red and blue regions in Fig.~\ref{fig:magneticphasediag}. The black color corresponds to the non-magnetic region. In Fig.~\ref{fig:anisotropies}(b) a wide range of values  ($0.2 < D_x/D_y < 1.7$) is found depending on the value of the interacting parameters. The experimentally observed transport anisotropy for undoped and electron-doped compounds~\cite{mazin10,chuscience2010,yingprl11,chuscience2012,fisherprl14} corresponds to $D_x/D_y > 1$.  These values coincide with the smallest values of the magnetic moment and appear within a great part of the LM and in the HM phases restricted to the nesting-like region and to the orbital differentiated region close to the crossover to the nesting-like in Fig.~\ref{fig:magneticphasediag}. This is believed to be the physical region for undoped iron arsenides, see Sec.~\ref{subsec:hund}. The result is in accordance with experiments in undoped and electron-doped samples since the resistivity anisotropy is largest close to the magnetic transition with small magnetic moment. 

$D_x/D_y > 1$ appears when the orbital ordering is small ($|n_{yz} −- n_{zx}| < 0.05$) and $D_x/D_y$ decreases with increasing orbital ordering in most part of the orbital differenciated region of the phase diagram. Thus, within this approximation, orbital ordering cannot be the origing of the transport anisotropy since it seems to favour the opposite anisotropy. This result is in agreement with other works~\cite{chen_devereaux10,daghofer09}. 
Deep in the strong orbital differentiation region, $D_x/D_y<1$ is related to the itinerancy of the $zx$, $x^2 −- y^2$, and $3z^2 −- r^2$ orbitals in the $y$-direction  which provides stability to the $(\pi,0)$ AF state (see Sec.~\ref{subsec:hund})~\cite{nosotrasprb12-2}.

Small orbital ordering and magnetic moments in the HM region with $D_x/D_y>1$ are consistent with reconstructed Fermi surfaces inferred from the ${\bf Q}=(\pi,0)$ doubling of the  antiferromagnetic unit cell. In Fig.~\ref{fig:bands} the reconstructed bands in the antiferromagnetic phase are shown at low energy. The magnetic bands are calculated for $U=1.6$ eV and $J=0.25U$ corresponding to a magnetic moment of $\sim 0.9\mu_B$ appropriate for the experimental values of 122 compounds. An orbital splitting of the $zx$/$yz$ weight at the $\Gamma$ pocket can be appreciated as observed in ARPES~\cite{shimojima10}. The connection of the $\Gamma$ pocket with the $X$ pockets by the ${\bf Q} = (\pi, 0)$ antiferromagnetic vector gives rise to the metallic nodal spin density wave~\cite{dunghailee09} with the formation of two Dirac cones along $\Gamma \rightarrow X$. These Dirac pockets govern the anisotropy~\cite{nosotrasprl10-2} in the yellow region of Fig.~\ref{fig:anisotropies}(b). In the orange region with the highest anisotropy additional small hole pockets appear around the $Y$ point of the Fermi surface (Lifsthiz transition). The particular form in which the Lifsthiz transition arises depends sensitively on the low energy physics of each tight-binding model but it will nevertheless influence the Drude anisotropy. This situation can be inferred from Fig.~\ref{fig:bands} shifting upwards the upper folded band with $zx$ component close to $Y$. 
The ratio value $D_x/D_y$ also depends on the shape of the Fermi surface. Therefore in this scenario, so-called "Fermi surface anisotropy", the anisotropy is strongly influenced by the topology and morphology of the Fermi surface. 
\begin{figure}
\includegraphics[clip,width=0.5\textwidth]{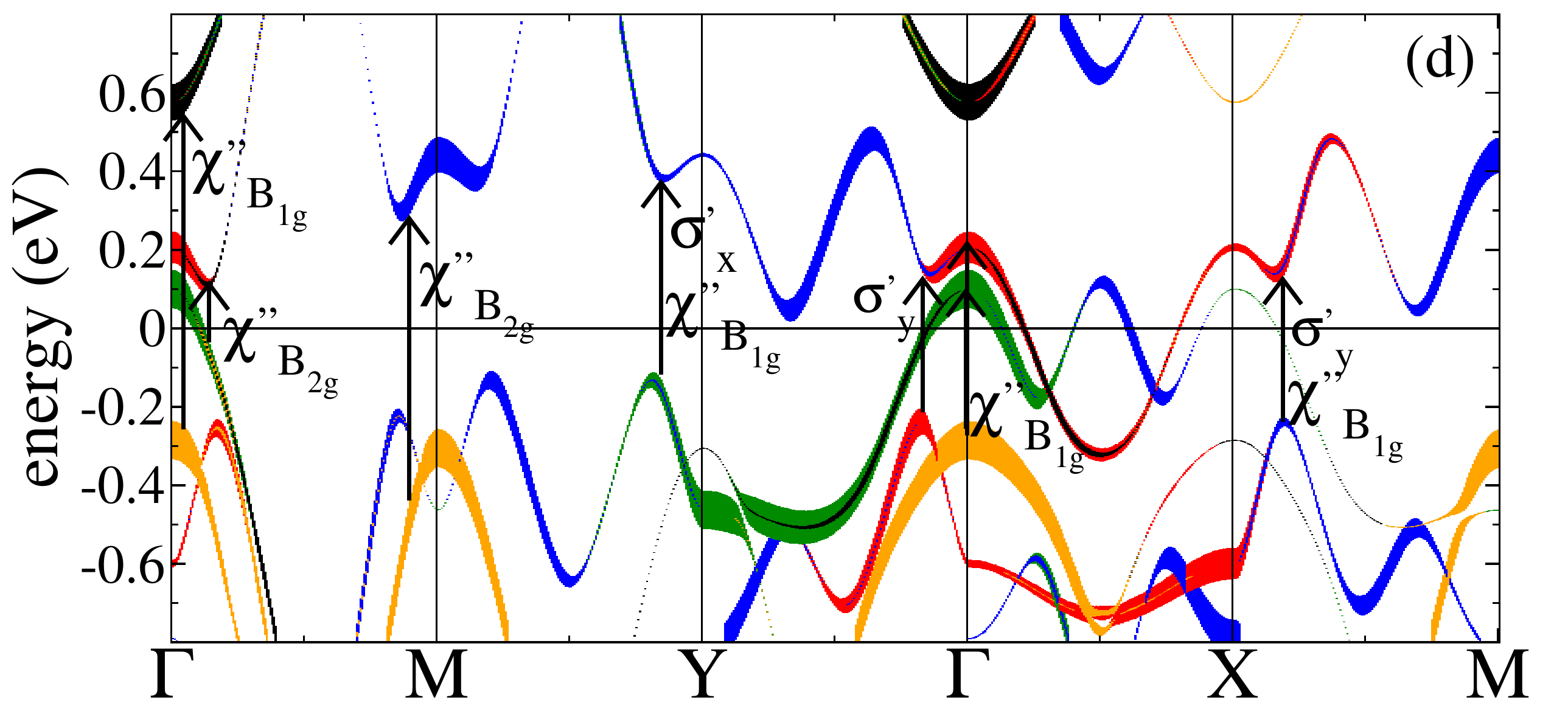}
\caption{(Color online) Energy bands in the $(\pi,0)$ antiferromagnetic state close to the Fermi level for $U=1.6$ eV and $J_H=0.25 U$. Linewidths and colors reflect the orbital content $yz$=red, $zx$=green, $xy$=blue, $3z^2-r^2$=orange and $x^2-y^2$=black. Note that not all the orbital content is visible due to overlapping curves. The optical ($\sigma_{xx}$, $\sigma_{yy}$) and Raman transitions ($\chi^{''}_{B_{1g}}$, $\chi^{''}_{B_{2g}}$) are shown. The figure is taken from Ref.~\cite{nosotrasprb13}.}
\label{fig:bands}
\end{figure} 
Other multi-orbital Hubbard models~\cite{japoneses11,dagottoprb11} and effective spin-fermion models~\cite{liangprl12,liangprl13} for appropriate values of the interactions and ab-initio calculations~\cite{mazin10}, have also found that the Fermi surface anisotropy corresponds to the experimental anisotropy. 

In Ref.~\cite{sugimotoprb14} the in-plane resistivity and the Drude weight were calculated in the antiferromagnetic state adding non-magnetic impurities to a five-orbital model. It was concluded that the anisotropy is enhanced by the effect of the impurity scattering. The scattering rate is strongly influenced by the change of the topology of the Fermi suface while the Drude weight is influenced by the topology and morphology of the Fermi surface in accordance to Ref.~\cite{nosotrasprl10-2}. 
Lifsthiz transitions driven by temperature in the reconstructed antiferromagnetic Fermi surface have also been claimed to be responsible for the anisotropy in the resistivity in the impurity scenario~\cite{hirschfeldarXiv14}. In summary, both scenarios, disorder driven anisotropy and Fermi surface anisotropy, are probably relevant to understand the anisotropy in the magnetic state and strongly depend on the topology of the reconstructed Fermi surface.

The Drude anisotropy extends to finite frequencies as seen in optical conductivity experiments~\cite{degiorgi10,uchida2011,degiorgi2012,nakajimaprl12,prozorovnatcomm13,mirriprb14,degiorgiprb14}. Optical conductivity along the antiferromagnetic $\sigma_{xx}(\omega)$ and ferromagnetic $\sigma_{yy}(\omega)$ directions show different intensity and peak frequencies. The Raman spectrum in the magnetic state shows signatures and peaks at energies similar to those found in optical conductivity~\cite{gallaisprb11,gallais2014}. Calculations in the Hartree-Fock approach for magnetism show that the different frequencies at which $\sigma_{xx}$ and $\sigma_{yy}$ peak reflect the magnetic gaps at $Y$ and $X$ electron pockets respectively. On the other hand, both interband transitions are active for the ${B_{1g}}$ polarisation of Raman experiments (see Fig.~\ref{fig:bands}). These transitions are based in the velocity and Raman vertices studied in Ref.~\cite{nosotrasprb13} and depend on the symmetry of the orbitals involved in the interband transitions. The dependence on $k$-space reflects the underlying lattice~\cite{nosotrasprb13}.
 This observation can help interpret optical and Raman data in the magnetic state.

In the nematic state the situation is even more complex since going beyond mean field like descriptions in five-orbital models or ab-initio calculations is very computationally demanding.
Two reviews of the nematic state by Y. Gallais and I. Paul and by A. Boehmer,  J. Schmalian and C. Meingast are presented in this volume~\cite{gallais-issue,meingast-issue}.
In this context Landau approaches have been shown to be very useful to understand the interplay between the structural and magnetic degrees of freedom and to calculate the response functions~\cite{lorenzanaprl08,Fernandesprl10,chubukovprb10,canoprb12,lorenzanaprb11,brydonprb11,schmalianprb12,Fernandesreview12,fernandesnatphys14}. Within the spin-nematic scenario~\cite{schmalianprb12,prozorovnatcomm13,breitkreizPRB2014}, the resistivity anisotropy originates in the scattering rate affected by the anisotropic spin fluctuations. In the weak coupling limit the spin nematic phase is derived from a microscopic Hamiltonian with hole and electron pockets without 
structure in the orbital degree of freedom~\cite{schmalianprb12}. The results
crucially depend on the ellipticity of the electron-pockets vanishing for circular electron Fermi surfaces~\cite{schmalianprb12, chubukovprb10}. 
However, these Fermi pockets posses a non trivial orbital composition as shown in Fig.~\ref{fig:lattice-FS}.
The particular arrangement of the $yz$ and $zx$ orbitals arises because under a 
90 degree rotation the two orbitals transform as $|xz\rangle \rightarrow 
|yz\rangle$ and $|yz\rangle \rightarrow -|xz\rangle$ ~\cite{scalapinoarXiv08}. 
Consequence of this singular $C_4$ symmetry is a non trivial topology manifested in the vorticity two of the 
$\Gamma$ pocket~\cite{scalapinoarXiv08, dunghailee09, lautimmprb13}. 

Recently an effective action has been proposed for the magnetic instability 
derived from a multi-orbital Hamiltonian~\cite{lauraarXiv14}. The Landau coefficients depend on the 
orbital content, Hubbard and Hund's coupling. In this approach it is uncovered that the orbital degree of freedom changes the spin-nematic scenario in an essential way due to the connection between the $\Gamma$ pocket with vorticity 2 and the $yz/zx$ composition of the 
$X/Y$ electron pockets.  It is found that in the nematic state the $yz/zx$ orbital degeneracy is lifted due to spin fluctuations without the need of adding a phenomenological crystal field term as in previous studies~\cite{lvphillipsprb10,schmalianprb12,breitkreizPRB2014}). These results extend the mean field findings of orbital ordering in the magnetic phase to the nematic state~\cite{leeyinku09, yu09, nosotrasprl10, nosotrasprl10-2, daghofer10, yinpickettprb10, yin11}. It is also found that the spin fluctuations in the nematic phase present anisotropic momentum dependence $\chi_{(\pi,0)}({\bf q}) \sim \chi_{(\pi,0)}(0)+a_x q_x^2+a_y q_y^2$ without invoking ellipticity of the electron pockets. This anisotropy is necessary to understand neutron scattering experiments at low energy~\cite{dialloprb10, hinkovprb10, mcqueeneyprb12, ibukaphysC14, luscience14} and is consistent with interpretations using ab-initio calculations~\cite{hinkovprb10, ibukaphysC14}. 

The $C_4$ symmetry breaking in the $(\pi,0)$ antiferromagnetic state is also reflected in the anomalous behavior of specific phonons~\cite{canfieldprb08,chauviere09,akrapprb09,letaconprb09,zhang10,basovprb11,uchida2011,gallaisprb11,kimultrafast2012,kumar12,avigo13,mandalprb14,retiggprl05}.
Particularly interesting is the case of the $A_{1g}$ As-phonon resonance in the Raman spectroscopy upon entering the magnetic phase. The $A_{1g}$  As-phonon involves arsenide atoms vibrations along the $z$-axis. In the tetragonal phase the $A_{1g}$ As-phonon is active neither in $B_{1g}$ nor in $B_{2g}$.  $B_{1g}$ and $B_{2g}$ polarisations are sensitive to anisotropic responses along the Fe-Fe bond and Fe-As bond respectively. A strong signal, which cannot be explained by the orthorombic distortion alone, appears just in $B_{1g}$ Raman response~\cite{gallaisprb11,sugai2012} when undergoing the magneto-structural transition . In Ref.~\cite{noelprb13} the Raman response is evaluated in the non-magnetic and in the magnetic states. It is found that the magnetic phase is able to induce a Raman intensity in the $B_{1g}$ Raman polarisation due to the $C_4$ symmetry breaking but the intensity is very sensitive to the way in which the coupling between the electrons and the phonons is introduced. With similar symmetry arguments, a finite Te-phonon intensity in the $B_{2g}$ polarisation was predicted to appear in the double stripe magnetic state of FeTe~\cite{noelprb13}.

\section{Magnetic softness}
\label{sec:softness}

Although the ($\pi,0$) columnar magnetic ordering is the most common one in the iron pnictides, other magnetic orders arise in Fe superconductors. FeTe orders in a double stripe pattern~\cite{baoPRL2009} and the alkaline iron selenides A$_y$Fe$_{2-x}$Se$_2$ have block antiferromagnetism and a particular Fe vacancy arrangement~\cite{weiCPL2011,dagottoreview-selenides}. Hole-doped Ba(Fe$_{1-x}$Cr$_{x}$)As$_2$ shows ($\pi,\pi$) order for $x>0.3$~\cite{MartyPRB2011}. Ferromagnetic order is much less usual than antiferromagnetism in Fe superconductors but it has been observed in a few cases~\cite{pachmayrAC2015,pachmayr2014,luNatMat2014,sunIC2015,dongarXiv2015,caoJPCS2012,nandiPRB2014,bendelePRB2013,monniPRB2013}. 
[(Li$_{1-x}$Fe$_x$)OH]FeSe based~\cite{pachmayrAC2015,pachmayr2014,luNatMat2014,sunIC2015,dongarXiv2015} and EuFe$_2$As$_2$ based~\cite{caoJPCS2012,nandiPRB2014}  compounds show FM (or canted AFM) in the hydroxide or Eu layer coexisting with superconductivity in the FeSe or FeAs layer. FeTe has been observed to turn FM under pressure~\cite{bendelePRB2013,monniPRB2013}. This variety of magnetic patterns indicates the magnetic order in these systems is soft, namely, it can be affected by small changes in parameters.

The magnetic softness has been addressed in the different approaches used to study the magnetic state of these materials. In general, the magnetic ground state is  dependent on the values of the parameters, chemical composition and details of the structure~\cite{nosotrasprb12}. Understanding the magnetic softness is important for superconductivity in these systems as ($\pi,0$) spin fluctuations lead to $s^{+-}$ pairing symmetry~\cite{mazin08,mazinrev09} while $(\pi,\pi$) fluctuations suppress $s^{+-}$ (and induce a d-wave state)~\cite{fernandesPRL2013}. FM correlations are expected to destroy singlet superconductivity.

\subsection{Magnetic softness in undoped compounds}

Weak coupling approaches assume the magnetic order is associated with a Fermi surface instability. Most of the Fe based superconductors show Fermi surface nesting between the $\Gamma$ hole pockets and the $X$ or $Y$ electron pockets~\cite{johnston_review2010}, see Fig.~\ref{fig:lattice-FS}, that would be expected to lead to ($\pi,0$) order. The alkaline Fe selenides, with no hole pocket in $\Gamma$ and hence no ${\bf Q}=(\pi,0)$ nesting, do have a different (block AF) order~\cite{dagottoreview-selenides}. However, some systems have a $(\pi,0)$ nested Fermi surface but do not show a $(\pi,0)$ magnetic ground state. This is the case of FeTe, which orders in a double stripe pattern~\cite{baoPRL2009} [with a ${\bf Q}=\pm (\pi/2,\pm \pi/2)$] which cannot be explained by nesting, and the phosphides, which do not order magnetically at all~\cite{kamihara06,luNat2008}.

The magnetic ground state of a strong coupling model that includes the antiferromagnetic exchange to first $J_1$ and second  $J_2$ neighbours is ($\pi,0$) for $J_1/J_2<2$~\cite{yildirim08}, and $(\pi,\pi)$ for $J_1/J_2>2$. In order to get the double stripe pattern, the inclusion of exchange up to third neighbours $J_3$ is required because the number of antiferromagnetic and ferromagnetic links is equal for both first and second neighbours~\cite{johannes09}. It has been argued that the stability of the double stripe order requires a very small $J_1$~\cite{ducatmanPRL2012} (which is a likely scenario for relatively large values of $J_H$, see Fig.~\ref{fig:exchange}) or the addition of a biquadratic term $K$~\cite{huPRB2012,valenti2015}.

\begin{figure}
\leavevmode
\includegraphics[clip,width=0.45\textwidth]{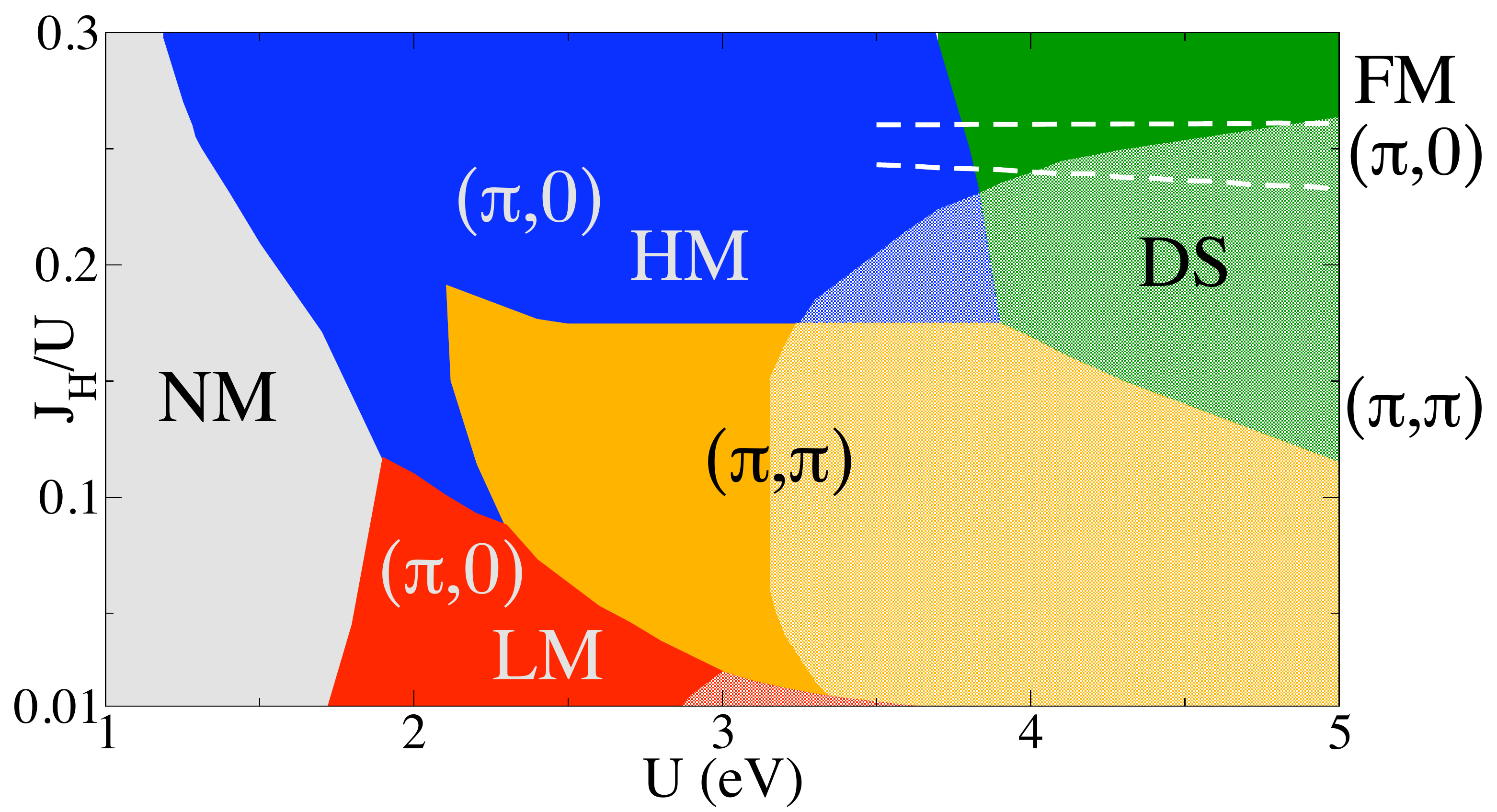}
\caption{Hartree-Fock phase diagram of a five-orbital model~\cite{nosotrasprb09} as a function of the onsite interactions $U$ and $J_H$ for undoped ($n=6$) Fe pnictides. The interorbital Coulomb repulsion $U'$ is given by $U'=U-2J_H$. Four different magnetic states have been considered. NM stands for non-magnetic, LM for low magnetic moment with violation of Hund's rule (antiparallel orbital magnetic moments), and HM for a magnetic state with parallel orbital magnetic moments. The ($\pi,\pi$) and double stripe regions (DS) also have a large value of the magnetic moment. Ferromagnetism (FM) is never the ground state for the parameters considered. Shaded areas correspond to insulating ground states (gap at the Fermi energy). Large $J_H/U$ promotes metallicity. The dashed lines and the labels on the right correspond to the results of calculating the energies within a Heisenberg Hamiltonian for $S=2$ spins with parameters calculated in second order perturbation theory for $t/U < 1$.}
\label{fig:PD-n6} 
\end{figure}

The relative success of the strong coupling approach to find the magnetic variability does not imply this model captures the full phenomenology of the Fe based superconductors, as argued in Sec.~\ref{sec:pi0}. For instance, although the double stripe order of FeTe could be explained in terms of localised spins~\cite{valenti2015}, itinerant electrons were invoked to explain the change in the effective spin per Fe from $S \sim 1$ in the antiferromagnetic state to $S\sim 3/2$ in the paramagnetic state revealed by an inelastic neutron scattering study of spin excitations~\cite{zaliznyakPRL2011}. Similarly, the transition to a magnetic spiral with increasing $y$ in Fe$_{1+y}$Te~\cite{baoPRL2009,rodriguezPRB2011} could be explained by a modification of the exchange couplings through the RKKY interaction between localised and delocalised electrons~\cite{ducatmanPRB2014}.
In general, models with localised and delocalised electrons have been proposed to explain the different magnetic phases as the competition between exchange couplings and kinetic energy driven order~\cite{yin10,lvphillipsprb10,nosotrasprb12-2,dagottonat12,rinconPRL2014}.

Magnetic softness is also patent in ab-initio calculations: different magnetic orders, with relatively close energies, can be stabilised.  Details in the chemical composition and the structure may lead to different ground states, which are often associated to different magnitudes of the magnetic moment. The ground state for the pnictides is the $(\pi,0)$ order, while for Fe$_{1+y}$Te it is a double stripe very closely followed in energy by a staggered dimer configuration~\cite{valenti2015} and $(\pi,0)$~\cite{johannes09}. This proximity in energy is consistent with the coexistence of magnetic scattering at both ${\bf Q}=(\pi,0)$ and double stripe momenta reported in Fe$_{1.02}$Te$_{1-x}$Se$_x$~\cite{liuNatMat2010}. FeSe, which experimentally does not show a magnetic transition unless high pressures are applied~\cite{bendelePRL2010}, has a staggered dimer ground state within ab-initio calculations~\cite{caoarXiv2014,valenti2015}. Ferromagnetic order is usually not stabilised in DFT calculations for undoped systems~\cite{johannes09},  an exception being the 11~\cite{valenti2015} and 111 compounds. Within DFT, in LiFeAs (which experimentally is a non-magnetic superconductor), both AFM and FM orders are possible, with AFM the ab-initio ground state. Calculations find a FM ground state in MgFeGe~\cite{jeschkePRB2013} and ferromagnetism in the related CuFeSb has been found in magnetisation and neutron scattering measurements~\cite{qianPRB2012}. These different magnetic orders -- $(\pi,0)$, double stripe and FM -- have been found to be the successive ground states for increasing values of the out-of-plane anion (the chalcogen or pnictogen) heights $z$. This relation has been explained from the different effect of $z$ on the values of the exchange parameters, with $J_3$ getting stronger and $J_{1,2}$ weaker as $z$ increases, as calculated for FeSe$_x$Te$_{1-x}$~\cite{moonPRL10}.

\begin{figure}
\leavevmode
\includegraphics[clip,width=0.48\textwidth]{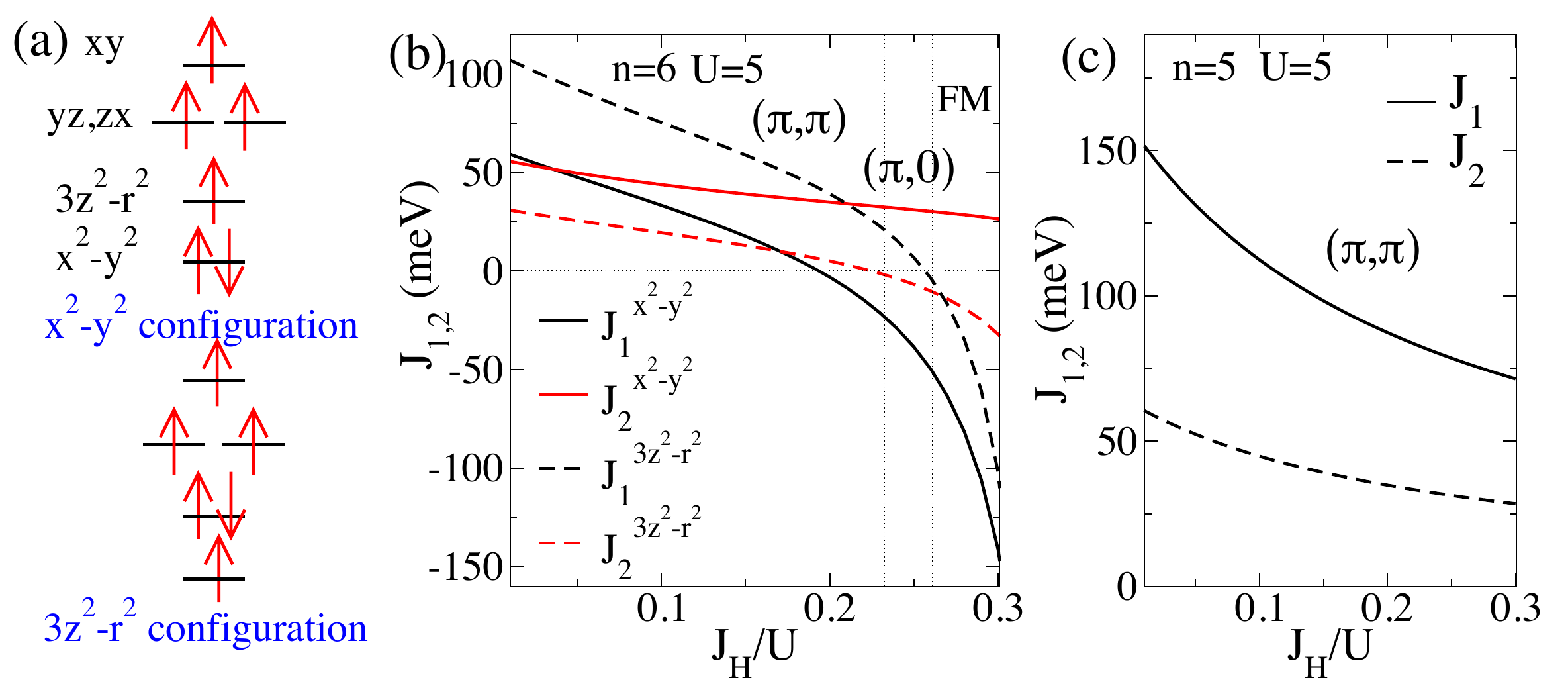}
\caption{Strong coupling approach to the Fe-superconductors phase diagram. The exchange couplings $J_1$ and $J_2$ are calculated within second order perturbation theory from the tight-binding model in Ref.~\cite{nosotrasprb09} assuming two different electron distributions among the orbitals as sketched in (a). The $x^2-y^2$ configuration considers this orbital to be doubly occupied. Similarly for the $3z^2-r^2$ configuration. (b) Exchange couplings to first $J_1$ and second $J_2$ neighbours for the two proposed configurations. The vertical dotted lines separate the regions in which ($\pi,\pi$), ($\pi,0$) and FM are the ground states for increasing values of $J_H/U$. (c) The exchange couplings calculated for half-filling ($n=5$) always give rise to a $(\pi,\pi)$ ground state.}
\label{fig:exchange} 
\end{figure}

The magnetic phase diagram of the undoped iron superconductors has been studied theoretically as a function of the interactions ($U$, and $J_H$) starting from different five orbital model hamiltonians, as explained in Sec.~\ref{sec:techniques}. 
Using a Hartree-Fock approximation in $k$-space for the undoped ($n=6$) iron-pnictides and comparing the energies of different magnetic orders [($\pi,0$), ($\pi,\pi$), double stripe, ferromagnetism, and non-magnetic ($m=0$)] for one of the models~\cite{nosotrasprb09} results in the magnetic phase diagram~\cite{nosotrasprb12} shown in Fig.~\ref{fig:PD-n6}. Ferromagnetism is never found to be the ground state and the double stripe order that appears for large values of the interactions is accompanied by charge ordering (but charge order is not observed experimentally). The appearance of the double stripe order for large values of the interactions is consistent with the finding that chalcogenides are more strongly correlated than pnictides~\cite{miyake2010}, see Sec.~\ref{sec:pi0}.

For intermediate values of the parameters, there is a close competition between the $(\pi,0)$ and ($\pi,\pi$) orders. This competition depends on the orbital filling (and it is hence related to orbital differentiation) and the crystal structure. A small variation of the crystal field can change the balance between these two orders: within Hartree-Fock, the transition between $(\pi,\pi)$ and $(\pi,0)$ is accompanied by a charge transfer from $3z^2-r^2$ to $x^2-y^2$~\cite{nosotrasprb12}.  Therefore, a slight shift down of the $x^2-y^2$ energy results in a more predominant $(\pi,0)$  region on the phase diagram~\cite{nosotrasprb12}. This softness is a consequence of the very small crystal field splitting between the orbitals. In general, small changes in the orbital filling among different models affect the ($\pi,\pi$)-($\pi,0$) phase diagram~\cite{brydonjpcm11}. For instance, the model in Ref.~\cite{graser09} has a relatively enlarged ($\pi,\pi$) region compared to the results shown in Fig.~\ref{fig:PD-n6}. 

The dependence of the ground state on the occupancy of the orbitals is also found in the strong coupling limit of the interacting models, see Eq.~\ref{eq:hamiltoniano}. Introducing the hopping as a second order perturbation to the interactions Hamiltonian, the five-orbital model can be mapped onto a Heisenberg model in which the couplings are calculated by adding the contributions from all the orbitals~\cite{nosotrasprb12}. The exchange couplings to first $J_1$ and second $J_2$ neighbours are calculated assuming that the orbitals are either singly or doubly (in the case of $x^2-y^2$ and $3z^2-r^2$) occupied (namely, the magnetic moment corresponds to $S=2$). $J_1$ and $J_2$ can be of the same order due to the indirect hoppings through the pnictogen or chalcogen~\cite{nosotrasprb09}, which connect both first and second neighbours.

Fig.~\ref{fig:exchange} shows the resulting $J_1$ and $J_2$ as a function of $J_H/U$. Two orbital configurations are distinguished: the $x^2-y^2$ configuration and the $3z^2-r^2$ configuration, whose name refers to which orbital is doubly occupied, see Fig.~\ref{fig:exchange}~(a). The values of the exchange couplings $J_{1,2}$ depend on the orbital configuration, the hoppings, and the interactions. The vertical lines in Fig.~\ref{fig:exchange}~(b) separate the regions with different magnetic order. These regions are superimposed with dashed lines in the Hartree-Fock phase diagram in Fig.~\ref{fig:PD-n6}. An increasing Hund's coupling $J_H$ suppresses both $J_1$ and $J_2$~\cite{haule09,nosotrasprb12}. $J_1$ decreases much faster than $J_2$, such that the condition for $(\pi,0)$ order ($J_1< 2J_2$) is fulfilled for a certain range of $J_H$ values. Consistent with Hartree-Fock results, the $(\pi,0)$ order corresponds to the $x^2-y^2$ configuration and the ($\pi,\pi$) order to the $3z^2-r^2$ configuration. This relation arises from the relatively large value of the $x^2-y^2$ hoppings between first neighbours~\cite{nosotrasprb09} and reveals the delicate balance between $J_H$ and the crystal field. The largest values of $J_H$ change the sign of the exchange couplings leading to FM order. The suppression of $J_1$ and $J_2$ is favoured by virtual transitions involving the doubly occupied $3z^2-r^2$ or $x^2-y^2$ orbitals~\cite{nosotrasprb12}.

$J_1$ and $J_2$ also depend on the angle $\alpha$ that characterises the Fe-As tetrahedra, see Fig.~\ref{fig:lattice-FS},  through the value of the hoppings~\cite{nosotrasprb09} leading to a predominance of $(\pi,0$) for squeezed tetrahedra and FM for elongated tetrahedra~\cite{nosotrasprb12}.

\subsection{Magnetic softness in doped compounds}

The magnetic softness also appears upon doping. 
At the extreme case of $n=5$ [Fig.~\ref{fig:exchange}~(c)], the strong coupling limit gives $J_1/J_2 > 2$ for all values of $J_H$ and the system has $(\pi,\pi)$ order. 
The competition between different magnetic orders as a function of the filling $n$ has been studied within the Hartree-Fock approximation in the reciprocal~\cite{nosotrasprb12} and in the real~\cite{luoPRB2014} spaces. The general trends of both sets of results are that ($\pi,\pi$) order is the ground state around $n=5$, which is expected at half-filling where correlations are stronger and the strong coupling result applies, and that the tendency towards FM increases with electron doping, with a $(\pi,0$) region around $n=6$, see Fig.~\ref{fig:PD-vs-n}. Phase separation appears in different parts of the phase diagrams. Many-variable variational Monte Carlo calculations for ab-initio low-energy models find the same qualitative features as a function of doping~\cite{imadaPRL2012}.

These theoretical results are consistent with measurements on materials in which Fe is substituted by a neighbouring transition metal while keeping the same crystal structure. For instance, Mn$^{2+}$ ions are half-filled ($d^5$) while Co$^{2+}$ has $7$ electrons ($d^7$). In BaMn$_2$As$_2$~\cite{singhPRB2009} or BaMnPnF (Pn=As, Sb, Bi)~\cite{saparov-SciRep2013}, a $(\pi,\pi)$ AF, insulating phase is found with very high $T_N$ and magnetic moment. On the electron-doped side of the phase diagram, LaCo(As,P)O are itinerant ferromagnets~\cite{sefatPRB2008}, CaCo$_2$As$_2$ has FM planes~\cite{anandPRB2014}, while BaCo$_2$As$_2$ is paramagnetic but allegedly close to ferromagnetism~\cite{sefatPRB2009}. From this, we should expect a tendency towards ($\pi,\pi$) order when doping with holes and towards ferromagnetism when doping with electrons. Consistently, neutron diffraction experiments~\cite{MartyPRB2011} in Cr ($d^4$) doped samples Ba(Fe$_{1-x}$Cr$_x$)As$_2$ show a transition between the ($\pi,0$) and the ($\pi,\pi$) order for $x \sim 0.3$. Note that partially substituting Fe by Cr or Mn does not produce a transition to a superconducting phase~\cite{thalerPRB2011}. It is believed that Cr and Mn do not incorporate itinerant electrons in the samples but act as magnetic impurities.

\begin{figure}
\leavevmode
\includegraphics[clip,width=0.48\textwidth]{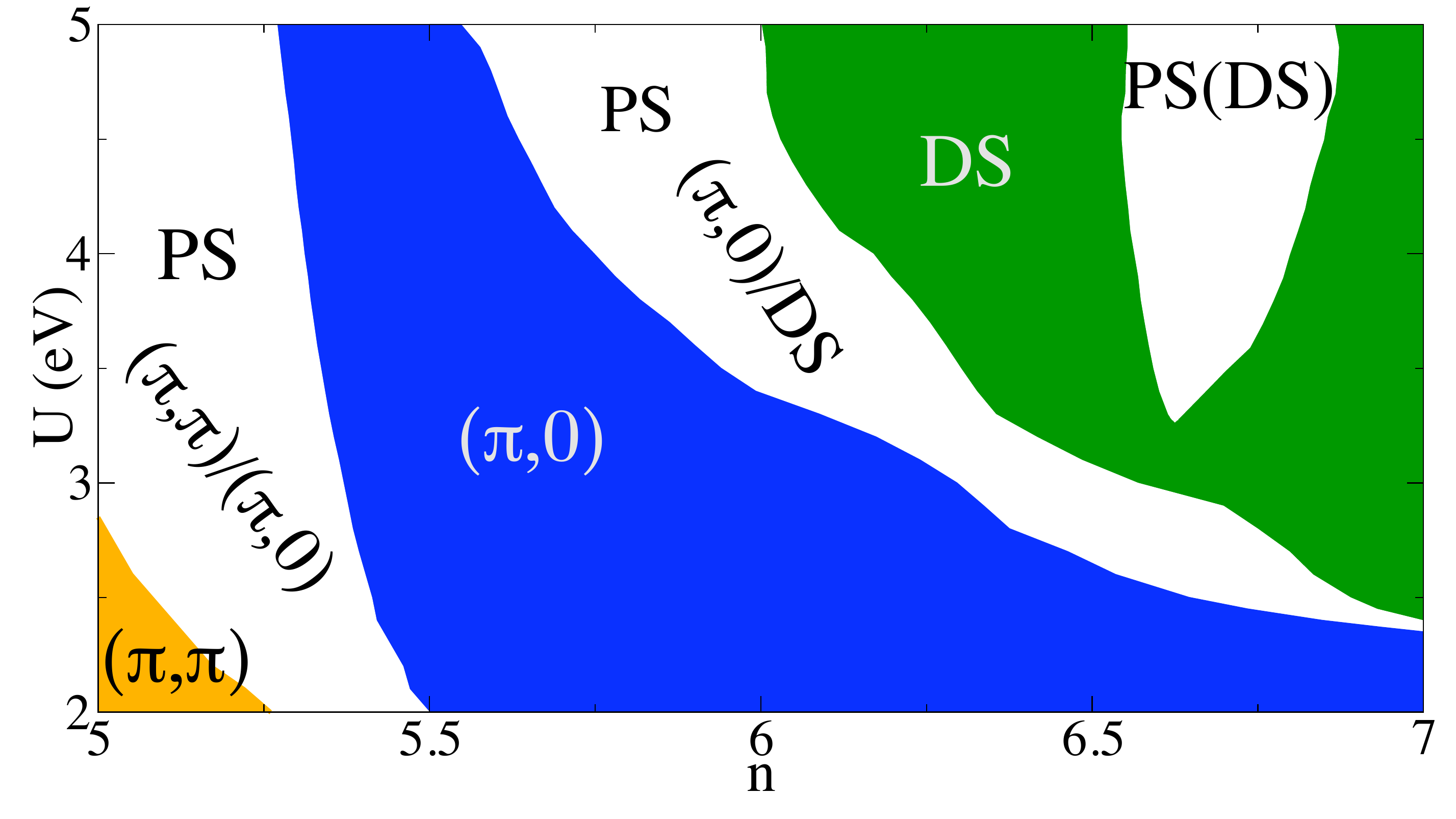}
\caption{Hartree-Fock phase diagram as a function of doping with the same multi-orbital model as considered in Fig.~\ref{fig:PD-n6}. The possible magnetic orders considered are $(\pi,0)$, $(\pi,\pi)$, double stripe and FM. The value of the Hund's coupling is fixed to $J_H=0.22U$. Hole doping ($n<6$) promotes antiferromagnetic interaction and the $(\pi,\pi)$ order is found close to $n=5$. Electron doping $n>6$ promotes the double stripe phase. There are big regions of phase separation in the phase diagram. }
\label{fig:PD-vs-n} 
\end{figure}

Neutron and X-ray diffraction measurements  on Ba(Fe$_{1-x}$Mn$_x$)$_2$As$_2$ reported the suppression of the orthorhombic to tetragonal structural transition at $x\sim0.1$ but that the Bragg peaks corresponding to stripe order [$(\pi,0)$ or $(0,\pi)$] survive in the tetragonal lattice for $x>0.1$~\cite{kimPRB2010}.  
Inelastic neutron scattering experiments~\cite{tuckerPRB2012} reported the coexistence of both ($\pi,0$) and ($\pi,\pi$) spin fluctuations  for Ba(Fe$_{0.925}$Mn$_{0.075}$)$_2$As$_2$. Fluctuations at ($\pi,0$) have a strong temperature dependence with its signal becoming broader above T$_N$. In contrast,  ($\pi,\pi$) fluctuations are very short ranged and weakly $T$ dependent, they are not suppressed when the ($\pi,0$) long range order sets in, and persist up to room temperature~\cite{tuckerPRB2012}. T$_N$ is suppressed upon Mn substitution up to $x \sim 0.1$ and it is enhanced for larger $x$ while the transition is significantly broadened~\cite{kimPRB2010,thalerPRB2011}. However, neutron Larmor diffraction together with nuclear magnetic resonance, muon spin-relaxation and neutron diffraction experiments~\cite{inosovPRB2013} for $x>0.1$ samples point to the coexistence of $(\pi,0)$ ordered orthorhombic areas and paramagnetic tetragonal areas. Antiferromagnetic short range ordered regions are seen to form at temperatures much higher than $T_N$~\cite{inosovPRB2013}. This interpretation as a coexistence phase is consistent with a model calculation that includes the interaction between the conduction electrons and the Mn magnetic impurities~\cite{gastiasoroPRL2014}.

A tetragonal (C$_4$ symmetry) phase has also been seen in various 122 systems close to the suppression of the antiferromagnetic phase, where there is a magnetism-superconductivity coexistence region. So far three compounds have shown this reentrant tetragonal phase: Ba$_{1-x}$Na$_x$Fe$_2$As$_2$~\cite{avciNatComm2014},  Ba$_{1-x}$K$_x$Fe$_2$As$_2$ ~\cite{bohmer2014}, and Sr$_{1-x}$Na$_x$Fe$_2$As$_2$~\cite{taddeiMM}. Ba$_{1-x}$Na$_x$Fe$_2$As$_2$ shows a C$_4$ phase coexistent with the orthorhombic one just at the antiferromagnetic transition suppression for a small range of doping~\cite{avciNatComm2014},  while this tetragonal region is fully inside the C$_2$ one in Sr$_{1-x}$Na$_x$Fe$_2$As$_2$~\cite{taddeiMM}.  The tetragonal phase in Ba$_{1-x}$K$_x$Fe$_2$As$_2$ is also at the magnetic transition but, differently to the other two cases, it is also reentrant as a function of temperature, being suppressed as the superconductivity sets in~\cite{bohmer2014}. The reentrant phase is accompanied by a spin reorientation~\cite{avciNatComm2014}. Polarised neutron diffraction data in single crystal samples of Ba$_{1-x}$Na$_x$Fe$_2$As$_2$ have identified the spin reorientation as occurring towards the out-of-plane c-axis~\cite{wasser2014}.

The magnetic order in the reentrant tetragonal C$_4$ phase is a matter of active research at the moment.
One possibility, consistent with a nesting origin of magnetism is the so-called double-Q order~\cite{lorenzanaprl08,ereminPRB2010,kimPRB2010,avciNatComm2014,wangPRB2014,kang2014,gastiasoroarxiv2015}. The double-Q order is a combination of $(\pi,0)$ and $(0,\pi)$ which fulfils the C$_4$ symmetry. This phase has been found within weak coupling models~\cite{ereminPRB2010}, on a two orbital~\cite{lorenzanaprl08} and  on a five orbital interacting model without disorder~\cite{gastiasoroarxiv2015}  within unrestricted Hartree-Fock, and, for the case of Mn-doped BaFe$_2$As$_2$, it has been argued via Ginzburg-Landau calculations to arise as the ground state between the $(\pi,0)$ C$_2$ phase and the $(\pi,\pi)$ order expected when the local moments on Mn dominate~\cite{wangPRB2014}. However, different magnetic and orbital order configurations may be compatible with the C$_4$ structure. It has been argued that a full characterisation of this phase could gives clues on the determination of the magnetic or orbital origin of the magnetoestructural transition~\cite{khalyavinPRB2014}.

\section{Summary and Conclusions}
\label{sec:conclusions}

The magnetic phase diagram of iron superconductors gives us information about the underlying electronic correlations in these systems. Such information may be crucial to understand the origin of the high T$_c$ superconductivity.
Here we have reviewed different proposals to explain the magnetic interactions in iron superconductors. The multiorbital character of these systems underlies their complex phenomenology and must be considered for the correct description of the electronic properties and the comparison with experiments. The magnetic interactions are directly linked to the strength of correlations in these materials, with Hund's coupling playing a very important role. Taking into account the orbital degree of freedom is crucial (a) to unify,  in a single $U$-$J_H$ phase diagram, the main mechanisms proposed to explain the $(\pi,0)$ magnetic order; (b) to explain the magnetic softness; (c) to address doping dependent properties; and (d) to understand the anisotropies.

Three different regions are  distinguished in the quasiparticle weight phase diagram of iron pnictides  plotted in Fig.~\ref{fig:hundmetal}: (i) a metallic region with $Z_\gamma \geq 0.5$ for moderate interactions, (ii)  a strongly correlated Hund metal with small quasiparticle weight, spin-polarized local atomic states, and orbital differentiation (i.e., orbital dependent correlations), for intermediate $U$ and $J_H$,  and (iii) a Mott insulating state at large $U$. The metallic regions (i) and (ii) are separated by a crossover. The interaction $U^*_{\rm Hund}$ at which this crossover takes place decreases in hole-doped materials, as the occupation of the Fe-d orbitals approaches half-filling.  

We argue that all the Fe-superconductors can be placed within this unified phase diagram. 
While the strength of correlations is still under debate, theoretical estimates and comparison with experiments suggest that the iron arsenides have $J_H/U \sim 0.15-0.3$, are not far from the crossover between the two metallic states, and become more correlated with hole-doping. Iron chalcogenides are believed to be more correlated, but still in the Hund metal region, or,  in the case of the alkaline-doped chalcogenides, close to the Mott insulating state.

The different states which appear in the $(\pi,0)$ Hartree-Fock phase diagram (Fig.~\ref{fig:magneticphasediag}) can be related to those in the correlations phase diagram (Fig.~\ref{fig:hundmetal}), and to weakly and strongly coupling descriptions proposed within different approaches, see discussion in Sec.~\ref{subsec:hund}. While Hartree-Fock cannot address the localisation of quasiparticles, the behavior and properties of the insulating state at large $U$ compares well with expectations from the strong coupling Heisenberg approach and the Mott insulating state. The magnetic region with strong orbital differentiation in Fig.~\ref{fig:magneticphasediag} seems related to the Hund metal state in Fig.~\ref{fig:hundmetal}. In this magnetic region $zx$, $3z^2-r^2$ and $x^2-y^2$ are itinerant, while $yz$ and $xy$ are half-filled and gapped. The interplay between these two sets of orbitals suggests a double-exchange mechanism for magnetism with itinerant and localized orbitals. Finally, the metallic region with moderate correlations  in Fig.~\ref{fig:hundmetal} is connected with three regions in the $(\pi,0)$ magnetic phase diagram: a non-magnetic state at small $U$, a magnetic ordered state which violates Hund's coupling, and a nesting originated itinerant magnetic state. Within this $(\pi,0$) magnetic phase diagram, the Fe arsenides are most probably within the crossover between the itinerant and the orbital differentiated regions.

The multiorbital character of iron superconductors allows for different magnetic ground states to arise in undoped systems. For intermediate values of $U$ and $J_H$, $(\pi,0)$ and $(\pi,\pi)$ compete. This competition can be tuned with a slight change in the relative crystal field between $x^2-y^2$ and $3z^2-r^2$. Larger values of the interactions, specially Hund's coupling, stabilise the double stripe, which is precisely the magnetic order found in iron chalcogenides, with strong correlations. Hole doping enhances correlations taking the systems closer to a Mott insulator and a $(\pi,\pi)$ order, while electron-doping promotes itinerancy and a tendency towards ferromagnetism, in accordance with experimental observation. Different magnetic states can be stabilised through small changes in the crystal structure, like the pnictogen (or chalcogen) height. 

The $C_4$ symmetry breaking of the $(\pi,0)$ antiferromagnetic state splits the degeneracy of the $yz$ and $zx$ orbitals and gives rise to orbital ordering as corroborated in multiorbital models. Spontaneous generation of orbital ordering has not been found without magnetism in realistic five orbital models. These models also find that both orbital ordering and magnetic moment are anticorrelated with the Drude anisotropy in the magnetic state. The topology and the morphology of the Fermi surface affects strongly to the resistivity anisotropy in the magnetic state in presence and in absence of impurities. The tetragonal symmetry breaking of magnetism is also behind the anomalous Raman response of the $A_{1g}$ phonon when undergoing the magnetic transition. 

Effective models accounting for the orbital degree of freedom are necessary to understand the nematic phase. Spin fluctuations give rise to $d_{xz}$/ $d_{yz}$ orbital splitting in this phase. Spin and orbital fluctuations are present then in the nematic state even in the spin-nematic scenario. The anisotropic properties of the spin fluctuations in the nematic phase depend sensitively on the singular $C_4$ symmetry of the $yz$ and $zx$ orbitals.

Seven years of intense research on iron superconductors have given rise to an unparalleled development of new concepts involving correlations in multi-orbital systems. Understanding their implications is likely the key to elucidate the mechanism of high T$_c$ superconductivity in these compounds. Iron superconductors promise to keep theorists and experimentalists busy for years to come. 

{\em Acknowledgements. } We thank Q. Si, R. Yu, A. Liebsch, L. de' Medici and N. Lanat\'a for sending their data and giving permission to use them in Fig.~\ref{fig:hundmetal}. We acknowledge funding from Ministerio de Econom{\'i}a y Competitividad (Spain) via Grants No. FIS2011-29689, No. FIS2012-33521 and No.FIS2014-53219-P and from Fundaci\'on Ram\'on Areces.


\bibliography{pnictides}

\begin{thebibliography}{299}
\bibitem[{fil({\natexlab{b}})}]{fillingnote2}
\bibinfo{note}{In LaFePO the mass is higher for $E_g$ orbitals because the
  crystal field splitting puts the $t_{2g}$ orbitals below the $e_g$ orbitals,
  thereby promoting a ground state with configuration $e_g^2t_{2g}^4$ in the
  atomic limit, see [\onlinecite{ferberprl2012}]}.
\end{thebibliography}

\end{document}